\documentclass[preprint,11pt]{elsarticle}

\usepackage{epsfig}

\usepackage{amssymb}
\usepackage{amsthm}
\usepackage{amsmath}

\def\tsc#1{\csdef{#1}{\textsc{\lowercase{#1}}\xspace}}
\tsc{WGM}
\tsc{QE}
\tsc{EP}
\tsc{PMS}
\tsc{BEC}
\tsc{DE}

\usepackage{geometry}
\usepackage{color,soul}
\usepackage{float}
\usepackage[caption = false]{subfig}

\usepackage[binary-units=true]{siunitx}
\DeclareSIUnit{\var}{Var}
\DeclareSIUnit{\va}{VA}
\DeclareSIUnit{\pu}{pu}
\DeclareSIUnit{\deg}{deg}

\usepackage{tikz}
\usetikzlibrary{positioning}
\usepackage{scalerel}
\usepackage{xcolor}
\usepackage{acronym}
\acrodef{vsc}[VSC]{Voltage Source Converter}
\acrodef{csc}[CSC]{Current Source Converter}
\acrodef{bess}[BESS]{Battery Energy Storage System}
\acrodef{pcr}[PCR]{Primary Control Reserve}
\acrodef{pll}[PLL]{Phase Locked Loop}
\acrodef{tso}[TSO]{Transmission System Operator}
\acrodef{gcp}[GCP]{Grid Coupling Point}
\acrodef{lso}[LSO]{Least Squares Optimisation}
\acrodef{pi}[PI]{Proportional--Integral}
\acrodef{pwm}[PWM]{Pulse Width Modulation}
\acrodef{mppt}[MPPT]{Maximum Power Point Tracking}
\acrodef{pmu}[PMU]{Phasor Measurement Units}
\acrodef{rms}[RMS]{Root Mean Square}
\acrodef{pss}[PSS]{Power System Stabilizers}

\journal{Sustainable Energy, Grids and Networks, published Sept. 2020}

\begin{document}

\begin{frontmatter}



\title{Frequency stability assessment of modern power systems: models definition and parameters identification\tnoteref{label2}}


\author[inst1]{Francesco Conte}

\affiliation[inst1]{organization={Dipartimento di Ingengneria Navale, Elettrica, Elettronica e delle Telecomunicazioni, University of Genova},
            city={Genova},
            country={Italy}}
            
\affiliation[inst2]{organization={Distributed Electrical Systems Laboratory, Ecole Polytechnique Fédérale de Lausanne (EPFL)},
            city={Lausanne},
            country={Switzerland}}

\author[inst1]{Stefano Massucco}  
\author[inst2]{Mario Paolone}
\author[inst1]{Giacomo-Piero Schiapparelli}
\author[inst1]{Federico Silvestro\corref{cor1}}
\author[inst2]{Yihui Zuo}

\cortext[cor1]{Corresponding Author mail: federico.silvestro@unige.it}
\fntext[label2]{DOI: https://doi.org/10.1016/j.segan.2020.100384}

\begin{abstract}
One of the fundamental concerns in the operation of modern power systems is the assessment of their frequency stability in case of inertia-reduction induced by the large share of power electronic interfaced resources. Within this context, the paper proposes a framework that, by making use of linear models of the frequency response of different types of power plants, including also grid--forming and grid--following converters, is capable to infer a numerically tractable dynamical model to be used in frequency stability assessment. Furthermore, the proposed framework makes use of models defined in a way such that their parameters can be inferred from real-time measurements feeding a classical least squares estimator.
The paper validates the proposed framework using a full-replica of the dynamical model of the IEEE 39 bus system simulated in a real--time platform.
\end{abstract}


\end{frontmatter}


\section{Introduction}

The recent technical literature has acknowledged the fact that modern and future power systems may experience reduction of their inertia due to the large-share of resources interfaced via power electronics. In this respect, system operators are, and will, facing new challenges to operate their grids safely, since phenomena that used to be uncommon in traditional power grids, such as large frequency modulations and rapid frequency variations, are more likely to be experienced.
In this context, the co-existence of conventional generation units with renewable resources needs to be included in the grid regulation schemes \cite{markovic2019understanding,MilanoPSCC2018,PSCC2018,TSE2019}.

The classical literature has clustered the global problem of power system stability in three main sub-categories: rotor angle stability, frequency stability and voltage stability \cite{kundur2004definition}. Recent works have discussed the effects of the integration of power electronic converters for each of these sub-categories, 
\emph{e.g.} small--signal stability of converter's controls \cite{yuan2017,rosso2019,rosso2019a}, current and voltage control loops and \ac{pll} \cite{goksu2014instability,rosso2019b,rosso2020,rokrok2019}; transient stability evaluations \cite{Wu2018pll_stability,farrokhabadi2020}; voltage stability under several fault conditions \cite{rosso2020,arani2016assessment,mortazavian2017dynamic,shabestary2016analytical}. 

Furthermore, frequency stability studies accounting for the presence of power electronic interfaces are mainly related to the influence of the active and reactive power control loops, which can be suitably designed to make the converter able to reproduce the behaviour of a synchronous machines \cite{zhang2016frequency,hammad2017effective,darco2015}.

In this respect, depending on their operation, power converters can be classified into \textit{grid--following} and \textit{grid-- forming}\footnote{Note that the recent literature has classified these converters using a slight different terminology \cite{rocabert2012control,green2007control,paquette2014,matevosyan2019}}. The former is based on a power converter whose injected currents are controlled with a specific phase displacement with respect to the grid voltage at the point of connection. As a consequence, the knowledge of the fundamental frequency phasor of the grid voltage at the point of connection is needed at any time for the correct calculation of the converter reference currents, whose amplitude and angle with respect to the grid voltage phasor are properly modified by outer control loops so as to inject the required amount of active and reactive power. The converter might entail additional outer loops modifying active and reactive current set-points, in order to provide regulation of grid frequency and grid voltage at the point of connection. This type of control does require the grid to be energised \cite{rocabert2012control}.

A grid--forming converter controls both frequency and voltage amplitude at its connection point providing either an a-static or droop-controlled frequency regulation. The grid-forming unit behaves always as a voltage source behind an impedance. As a consequence, the knowledge of the fundamental frequency phasor of the grid voltage at the point of connection is not needed. Since the current provided by the power converter is not necessary controlled, its output impedance can be modified according to the operating conditions in order to satisfy the current limitations of the converter and prevent from hardware damages. The classical literature has named these converters as \ac{vsc}. In view of the above, this type of converters does not require, in general, the use of a \acp{pll} \cite{rocabert2012control,matevosyan2019}. 

Most of the existing literature has addressed the problem of system stability usually considering converters operating in parallel with other similar devices \cite{rosso2019a,Wang2018,migrate3_2, wen2013influence} or connected to an infinite busbar \cite{rosso2019,rosso2019a,qoria2019power}.  Many studies are focused on the determination of the system stability using impedance-based criteria \cite{sun2011impedance, wang2014modeling}. Frequency stability studies are usually related to microgrids or isolated networks \cite{farrokhabadi2020,bottrell2013}, and power converters are represented with simplified models (\emph{e.g.} a simple delay) \cite{guo2014distributed, kerdphol2019self}, especially when the regulation is provided by a mixture of converter-based units and traditional generation. The mix of traditional units and power converters, operated with different control strategies such as grid--forming or grid--following, all cooperating to the frequency control, is not usually considered in literature, even if it represents the most probable scenario for the next future.

In this paper, we consider the problem of frequency stability in a scenario that involves the presence of traditional and converter--based units both providing frequency control.
The purpose is to define a numerically-tractable multi-area equivalent linear model of the original dynamical system, suitable to be used in frequency stability studies. The end--user of such a model is the system operator, who has availability of field measurements and desires to investigate frequency stability of his grid, using time-domain simulation and equivalent modeling in order to study new control strategies. 
Specifically, the defined multi-area linear model can be used for tuning and testing regulation strategies (\emph{e.g.} primary, secondary loops, \ac{pss}, $H_{\infty}$ robust control \cite{bevrani2014robust}), or for the definition of simpler model based on reduction techniques \cite{Gu2018}.
The proposed formulation integrates an algorithm to infer the model's parameter from the field measurements suitably coupled with a classical least squares estimator.

\begin{figure}
	\centering
	\includegraphics[width=0.6\columnwidth]{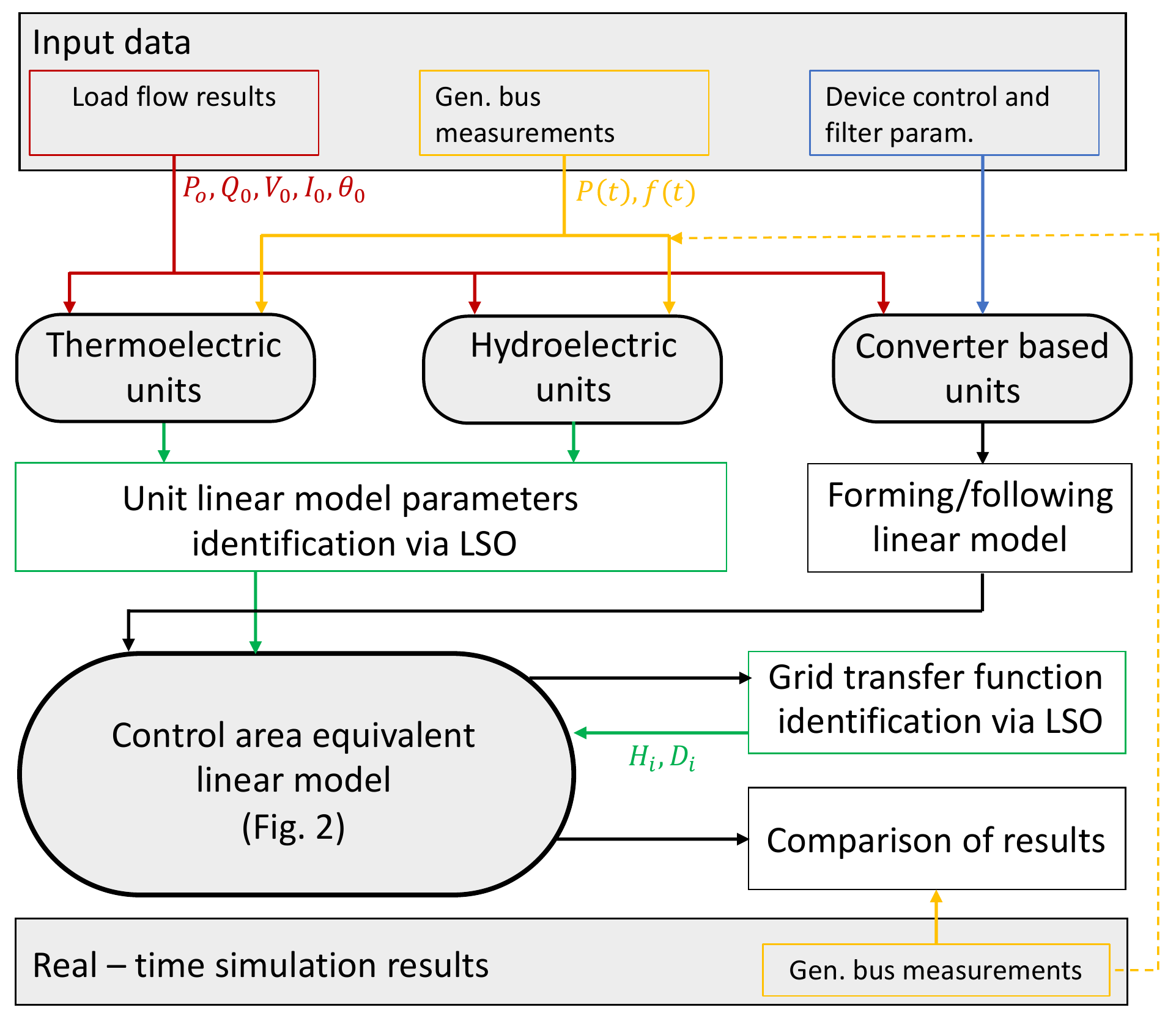}
	\caption{Flowchart describing the proposed approach for equivalent modeling and stability assessment.}\label{fig:flowchart}
\end{figure}

The flowchart in Figure~\ref{fig:flowchart} summarises the proposed approach. Specifically, the aim is to develop a linear model for each system control area which is potentially composed by both traditional thermal and/or hydroelectric units and converter based units. 
Classical models of generating units are adapted form classical literature and indetified with a \ac{lso} performed using power, $P(t)$ and frequency measurements, $f(t)$, as inputs. The converter based units models are obtained through the linearization of the system model and control loops. Models are tuned by using known control and system parameters. When not available, converters parameters can be optionally identified using \ac{lso}. Load flow results (such as active and reactive powers, $P_0$ and $Q_0$, voltages $V_0$, currents $I_0$ and angles $\theta_0$), define the initial operating point. 

The area equivalent model is tuned performing a second \ac{lso} which defines equivalent inertia $H_i$ and damping coefficients $D_i$. Finally, the proposed approach is validated by means of a comparison with the results obtained by a full-replica dynamical model of the IEEE 39 bus system implemented into a real--time simulator. 

The contributions of the paper are:
 \emph{(i)} development of an equivalent linear model for grid--following converters, \emph{(ii)} introduction of a rigorous approach to include grid--following and grid--forming units into linear models to be used in traditional frequency stability studies; \emph{(iii)}  definition of a strategy to identify the parameters of these equivalent models based on the knowledge of power and frequency measurements and the initial system equilibrium point. 
As mentioned, an original contribution of this paper is the integration of the equivalent linear model of grid--forming converters proposed in \cite{migrate3_2} with a suitably developed equivalent linear model of grid--following converters. This model is obtained through a rigorous linearization process, considering the full-replica of the original nonlinear model and applying usual hypothesis adopted in classical frequency stability studies (\textit{e.g.} constant voltage). The motivation behind the development of an equivalent linear model is to allow for its parameter identification, which can be carried out using standard \ac{lso} techniques designed for linear models.
 
We finally validate the approach with comparison with a non--linear dynamical model implemented on a real--time simulator. The validation and the numerical results also show the practical implementation of the procedure also for online identification of the linear models and online frequency stability assessment.

The rest of the paper is organised as follows: in Section~\ref{sec:Models} the linear model of the single resources for frequency stability studies is presented. Section~\ref{sec:Parameters_identification} describes the strategy for the identification of the parameters of the equivalent linear model using data coming from real--time measurements. In Section~\ref{sec:real_time_simulator} the real--time simulation setup used to validate the approach is illustrated. Section~\ref{sec:results} reports the validation results. Section~\ref{sec:Conclusions} concludes the paper with a summary of the main findings of this research.

\section{Models for frequency stability studies} \label{sec:Models}
The primary frequency control schemes, typically divide the power system into groups of generators and loads named control areas. A multi--area power systems consists of a set of areas interconnected by tie-lines.
Two classical hypotheses are adopted:
\begin{enumerate}
    \item frequency is equal within a control area \cite{bevrani2014robust};
    \item voltage phenomena are faster than the frequency dynamics and thus they can be neglected.
\end{enumerate}

\begin{figure}
	\centering
	\includegraphics[width=0.8\columnwidth]{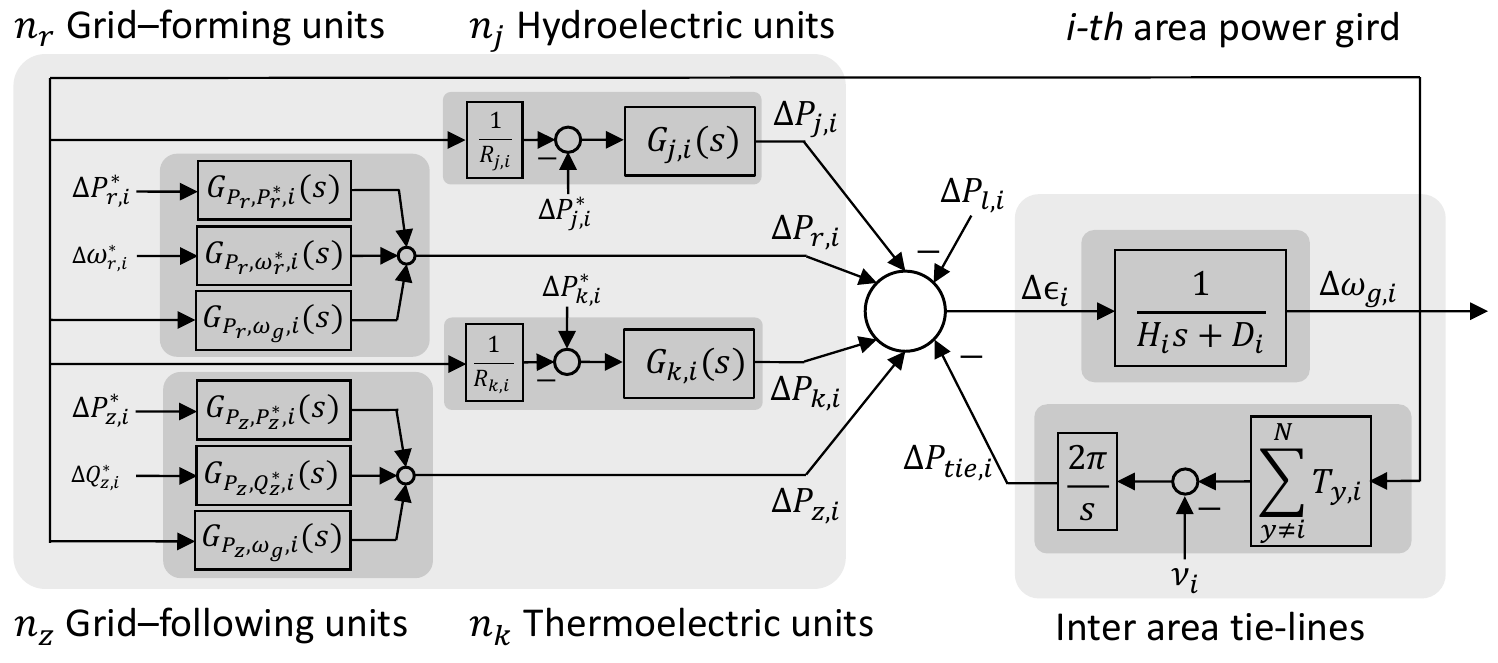}
	\caption{Control area block diagram for primary frequency control. }\label{fig:areamodel}
\end{figure}

In this paper, we consider the representation of a multi-area power system as illustrated by Fig.~\ref{fig:areamodel}. Specifically, the \emph{i-th} area is composed by $n_{j,i}$ hydroelectric units, $n_{k,i}$ thermoelectric units, $n_{r,i}$ grid--forming units and $n_{z,i}$ grid--following units each one modelled by one or more transfer functions. The area power system, and the associated inter--area connections, are also included. The details of each control block are provided in the following subsections.

\subsection{Converter--based power plants}
\begin{figure}
	\centering
\subfloat[ ]{\includegraphics[width=0.8\columnwidth]{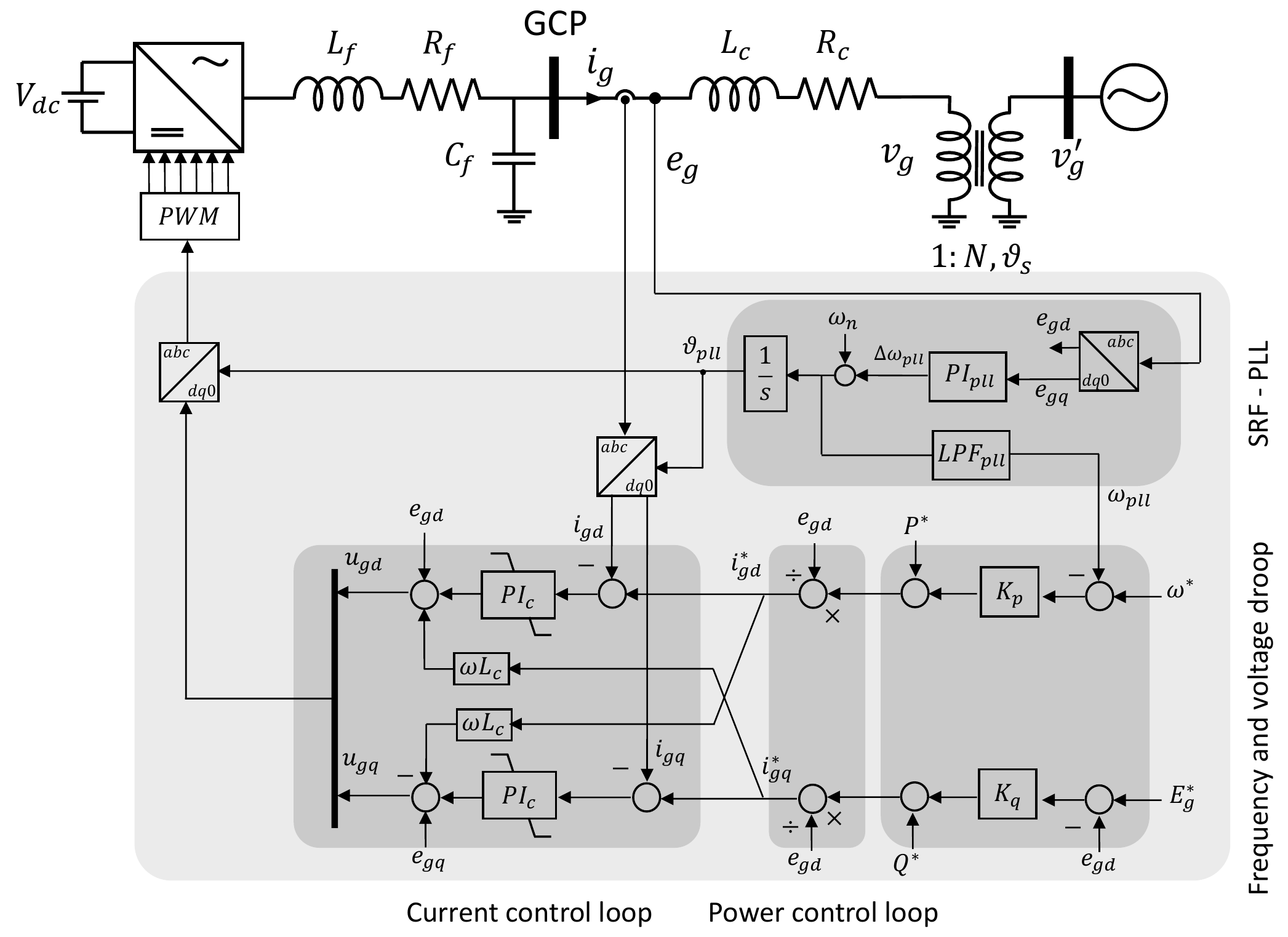}}\\
\subfloat[ ]{\includegraphics[width=0.8\columnwidth]{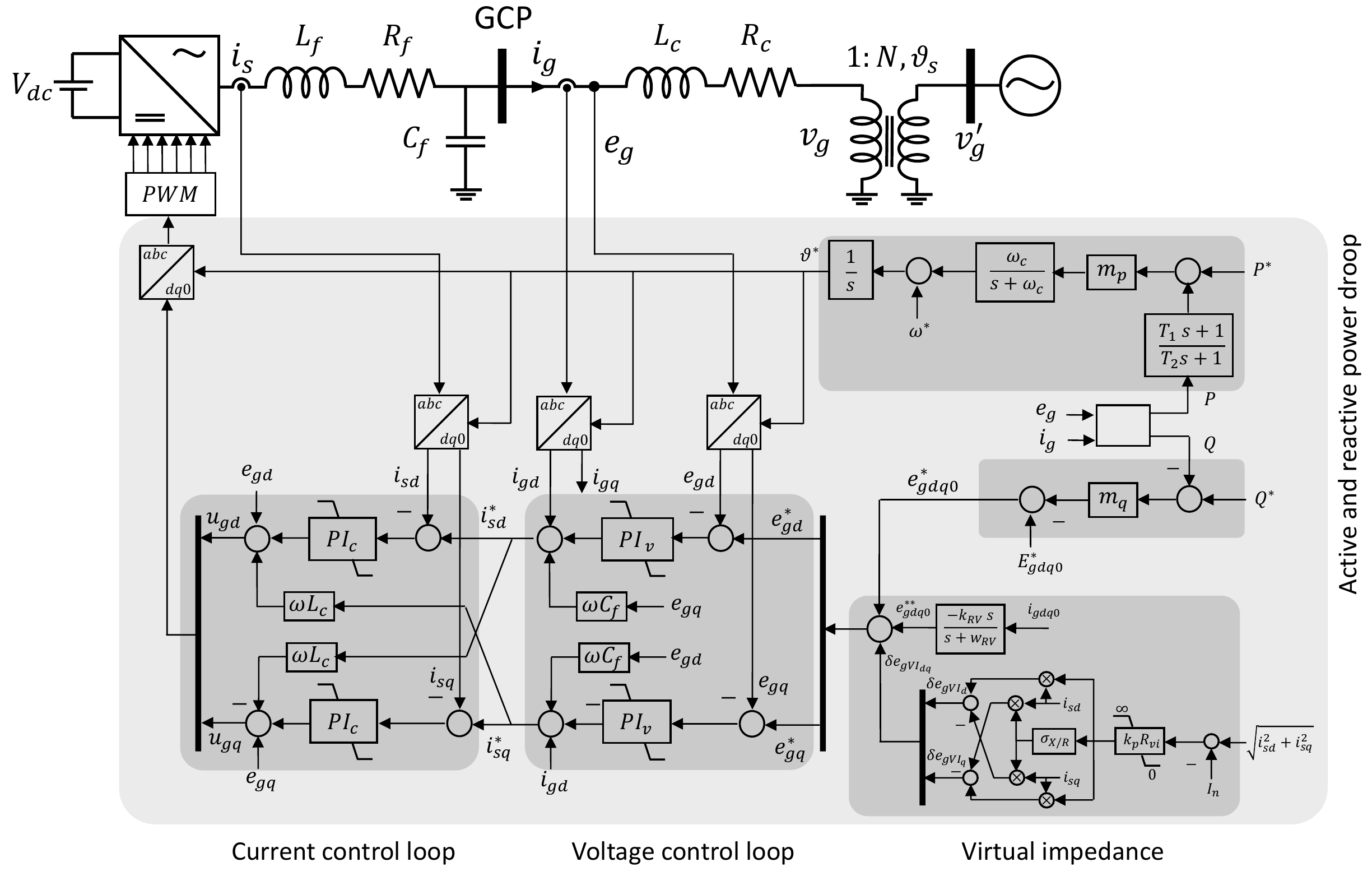}}
\caption{(a) Grid--following and (b) grid--forming converter both operating in grid-supporting mode.}
\label{fig:grid_forming_feeding_control_loop}
\end{figure}

Figure \ref{fig:grid_forming_feeding_control_loop} shows the schemes of the two classes of power converters considered in this paper: grid--following and grid--forming.
In general, power converters can be divided into three main components: switching devices, grid coupling filter and controls. Switching devices and grid coupling filters constitute the physical part of the converter and they are the same both for grid--following and grid--forming configurations. Control and filtering have a significant effect on the system performance \cite{green2007control}. The filtering action is usually provided by a third-order LCL filter, composed by the summation of the LC converter's filter and the coupling transformer or the coupling line inductance \cite{green2007control}.

The control that can operate the converter as a grid-- following or a grid--forming unit, presents an inner and an outer loop. The inner control loop, usually similar for both the converter types, is the \emph{-dq} current \ac{pi} controller, which defines the reference signal for the \ac{pwm}. Moreover, the performance of this control loop are improved by the inclusion of the grid voltage feed--forward and the decoupling network term \cite{timbus2009evaluation}.

The outer controllers are, instead, different for the grid--following and grid--forming converters. Indeed, the outer control of grid--following converters defines the current reference provided to the current loop from active and reactive power set--points (these last can be provided by higher-level controllers such as \ac{mppt}, dispatch or energy management systems). When the grid--following operates in grid--supporting mode, the power set--points present additional terms proportional to the frequency deviation and to the voltage deviation, respectively (Fig.~\ref{fig:grid_forming_feeding_control_loop}-a). The
frequency measurement, as well as the angle reference for the \emph{-dq} transformation, are given by the SFR--\ac{pll}.

The current references for the grid--forming converter are set by a voltage \ac{pi} control loop. Even in this case, a current feed--forward could be used to improve the control's performance. The voltage \emph{-dq} reference is given to the controller externally as well as the frequency reference which is also used to set the angle reference for the \emph{-dq} transformation.
The grid--forming converter can also operate in grid--supporting mode by setting the voltage reference through a virtual impedance controller used to emulate the behaviour of synchonous machine (Fig.~\ref{fig:grid_forming_feeding_control_loop}-b). Specifically, as in the simplified scheme of a synchronous generator, the converter voltage is linked to the grid with an impedance \cite{rocabert2012control,migrate3_2}. Moreover, the frequency reference is set by the summation of the nominal value and a term proportional to the power deviation in the so called reverse droop control. Additional filters can be added to this droop controller in order to improve stability and performance \cite{migrate3_2}. 

\subsubsection{Physical component model}
Let us consider the coupling transformer between the converter and the grid shown in Fig.~\ref{fig:grid_forming_feeding_control_loop}. The physical part of the converter is modelled by the following equations, expressed in per unit (\SI{}{\pu}) and in \emph{dq} coordinates:
\begin{align}   
    & \frac{L_c}{\omega_b} \frac{d i_{gd}}{dt} = e_{gd} - v_{gd} - R_c i_{gd} + \omega_g L_c i_{gq} \label{eq:syst_pu_d}\\
    & \frac{L_c}{\omega_b} \frac{d i_{gq}}{dt} = e_{gq} - v_{gq} - R_c i_{gq} - \omega_g L_c i_{gd} \label{eq:syst_pu_q}.
\end{align}
where $L_c$ and $R_c$ are in \SI{}{\pu} of the transformer's basis and $\omega_g =$ \SI{1}{\pu} is the nominal system frequency. The system base values assumed in this paper are: $S_b =$ \SI{100}{\mega\watt}, $V_{b1} = $\SI{22}{\kilo\volt}, $V_{b2} = $\SI{345}{\kilo\volt}, $Z_{bi} = V_{bi}^2/S_b$, $Y_{bi} = 1/Y_{bi}$, $f_b = $ \SI{50}{\hertz}, $\omega_b = 2 \pi f_b$ \si{\radian\per\second}.
Additionally, in model \eqref{eq:syst_pu_d}--\eqref{eq:syst_pu_q}, the voltage phasors result to be $\dot{v}_g = V_{g}e^{j\vartheta_g}=v_{gd} + j v_{gq}$ and $\dot{e}_g = E_{g}e^{j\delta} = e_{gd} + j e_{gq}$. Moreover, a fixed grid voltage  $\dot{v}_{g0} =V_{g0}e^{j\vartheta_{g0}} = v_{g0d} + j v_{g0q}$ is assumed.

 Model \eqref{eq:syst_pu_d}--\eqref{eq:syst_pu_q} is linearized to obtain the following multi-input multi-output (MIMO) transfer function: 
\begin{align}      
    \begin{bmatrix} \Delta i_{gd}(s)\\ \Delta i_{gq}(s) \end{bmatrix} =
    \begin{bmatrix} G_{i_d,\delta }(s) & G_{i_q,E_{g}}(s) \\ G_{i_q,\delta }(s) & G_{i_d,E_{g}}(s) \end{bmatrix} 
    \cdot \begin{bmatrix} \Delta \delta (s) \\\Delta E_{g} (s)  \end{bmatrix}  \label{eq:MIMO1}
\end{align}

Moreover, the converter active and reactive power at the \ac{gcp} expressed for the \emph{dq} frame are,  $P = i_{gd} e_{gd} + i_{gq} e_{gq}$ and  $Q = i_{gq} e_{gd} - i_{gd} e_{gq}$. The linearisation of these equations returns the following MIMO transfer function:
\begin{align}      
    \begin{bmatrix} \Delta P(s)\\ \Delta Q(s) \end{bmatrix} =
    \begin{bmatrix} T_{P,\delta }(s) & T_{P,E_{g}}(s) \\ T_{Q,\delta }(s) & T_{Q,E_{g}}(s) \end{bmatrix} 
    \cdot \begin{bmatrix} \Delta \delta (s) \\\Delta E_{g} (s)  \end{bmatrix} . \label{eq:MIMO2}
\end{align}

The expressions of the transfer functions in \eqref{eq:MIMO1} and \eqref{eq:MIMO2}, listed in the following, are computed using the linearizations
 \begin{align}  
     & e_{gd}  \approx E_{g0} \cos{\delta_0} - E_{g0} \sin{\delta_0} \Delta \delta +\cos{\delta_0} \Delta E_{g},  \label{eq:lin_init}\\
    & e_{gq}  \approx E_{g0} \sin{\delta_0} + E_{g0} \cos{\delta_0} \Delta \delta +\sin{\delta_0} \Delta E_{g}
    \end{align}
and the relations
\begin{align}
    & v_{gd} = V_{g0} \cos{\vartheta_{0}},\\
    & v_{gq} = V_{g0} \sin{\vartheta_{0}},  \label{eq:lin_end}
\end{align}
where the subscript $0$ indicates the operating points. 

For the MIMO transfer function \eqref{eq:MIMO1}, we have:
 \begin{align}      
 &   G_{i_d,\delta }(s)=   \frac{  -  \sin{\delta_0}    \left( \frac{L_c}{\omega_b}  s + R_c \right) +     \omega_g L_c  \cos{\delta_0} }{ \left( \frac{L_c}{\omega_b}  s + R_c \right)^2 + \omega_g^2 L_c^2 }  E_{g0},  \label{eq:G_id_delta} \\
    & G_{i_q,\delta }(s) =  \frac{\cos{\delta_0} \left( \frac{L_c}{\omega_b}  s + R_c \right)  +   \omega_g L_c \sin{\delta_0} }{ \left( \frac{L_c}{\omega_b}  s + R_c \right)^2 + \omega_g^2 L_c^2 }  E_{g0}.   \label{eq:G_iq_delta}
\end{align}
\begin{align}      
    & G_{i_d,E_{g}}(s)  =  \frac{ \cos{\delta_0}\left(\frac{L_c}{\omega_b} s + R_c\right) + \omega_g L_c \sin{\delta_0}   }{\left(\frac{L_c}{\omega_b} s + R_c\right)^2 +\omega_g ^2 L_c^2}    \label{eq:G_id_E},\\
        &   G_{i_q,E_{g}}(s) =  \frac{\sin{\delta_0}  \left(\frac{L_c}{\omega_b} s + R_c\right) -  \omega_g L_c  \cos{\delta_0} }{\left(\frac{L_c}{\omega_b} s + R_c\right)^2 +\omega_g ^2 L_c^2}   \label{eq:G_iq_E}, 
\end{align}

The MIMO transfer function \eqref{eq:MIMO2}, is obtained substituting     \eqref{eq:G_id_delta}--\eqref{eq:G_iq_E}  in the linearization of the powers at \ac{gcp},
\begin{align}      
& T_{P,\delta }(s)   = \frac{ \beta_{2,P\delta} s^2   +\beta_{1,P\delta} s + \beta_{0,P\delta} }{  \left( \frac{L_c^2}{\omega_b^2}\right)   s^2 +  \left( \frac{2  R_c L_c}{\omega_b}\right)  s + \left( R_c^2 +\omega_g^2 L_c^2\right)  }  \label{eq:G_P_delta} ,  \\
& \beta_{2,P\delta} = \frac{Q_0 L_c^2}{\omega_b^2}  \quad \quad \quad \beta_{1,P\delta} =    \frac{2 Q_0 R_c L_c}{\omega_b}  \nonumber, \\
&\beta_{0,P\delta} = Q_0 (R_c^2 +  \omega_g^2 L_c^2 ) +  E_{g0}^2  \omega_g L_c   \nonumber, \\
& T_{Q,\delta }(s)  =   \frac{ \beta_{2,Q\delta}  s^2 + \beta_{1,Q\delta} s  +
\beta_{0,Q\delta}   }{ \left( \frac{L_c^2}{\omega_b^2}\right)   s^2 +  \left( \frac{2  R_c L_c}{\omega_b}\right)  s + \left( R_c^2 + \omega_g^2 L_c^2\right) }   \label{eq:G_Q_delta}, \\
 & \beta_{2,Q\delta} = -\frac{ P_0 L_c^2}{\omega_b^2}  \quad \quad \quad \beta_{1,Q\delta} = E_{g0} ^2 \frac{L_c}{\omega_b}  s - \frac{ 2 P_0 L_c  R_c}{\omega_b}\nonumber, \\
 & \beta_{0,Q\delta} =  E_{g0} ^2 R_c - P_0  (R_c^2 + \omega_g^2 L_c^2)   \nonumber.
\end{align}
\begin{align}      
& T_{P,E_{g}}(s)  =  \frac{ \beta_{2,PE} s^2 + \beta_{1,PE} s +    \beta_{0,PE} }{\left(\frac{L_c^2}{\omega_b^2} \right)s^2+  \left(\frac{2 R_c L_c}{\omega_b} \right)s+ \left( R_c^2+\omega_g^2 L_c^2 \right)}   \label{eq:G_P_E} ,\\
    & \beta_{2,PE}= \frac{P_0}{E_{g0}} \frac{L_c^2}{\omega_b^2}  \quad  \beta_{1,PE}=   \frac{P_0}{E_{g0}} \frac{2 L_c R_c}{\omega_b} + E_{g0}\frac{L_c}{\omega_b} \nonumber ,\\
    &\beta_{0,PE}=  \frac{P_0}{E_{g0}} R_c^2 + \frac{P_0}{E_{g0}}\omega_g ^2 L_c^2 +E_{g0} R_c \nonumber, \\
    & T_{Q,E_{g}}(s)= \frac{ \beta_{2,QE} s^2 + \beta_{1,QE}s +\beta_{0,QE} }{\left(\frac{L_c^2}{\omega_b^2} \right)s^2+  \left(\frac{2 R_c L_c}{\omega_b} \right)s+ \left( R_c^2+\omega_g^2 L_c^2 \right)}  \label{eq:G_Q_E}, \\
    & \beta_{2,QE}=  \frac{Q_0}{E_{g0}} \frac{L_c^2}{\omega_b^2} \quad 
      \beta_{1,QE}=  \frac{Q_0}{E_{g0}} \frac{2 L_c R_c}{\omega_b} \nonumber, \\
    &\beta_{0,QE}=   \frac{Q_0}{E_{g0}}\left( R_c^2  +\omega_g ^2 L_c^2\right) - E_{g0} \omega_g L_c  .\nonumber 
\end{align}

\subsubsection{Grid--forming converter}
As discussed above, the grid--forming converter can be considered as an ideal voltage source with a low--output impedance, setting the voltage amplitude and frequency of the grid by using a proper control loop and, in general, it does not require the usage of \acp{pll} \cite{rocabert2012control}. 

\begin{figure}
	\centering
	\includegraphics[width=0.9\columnwidth]{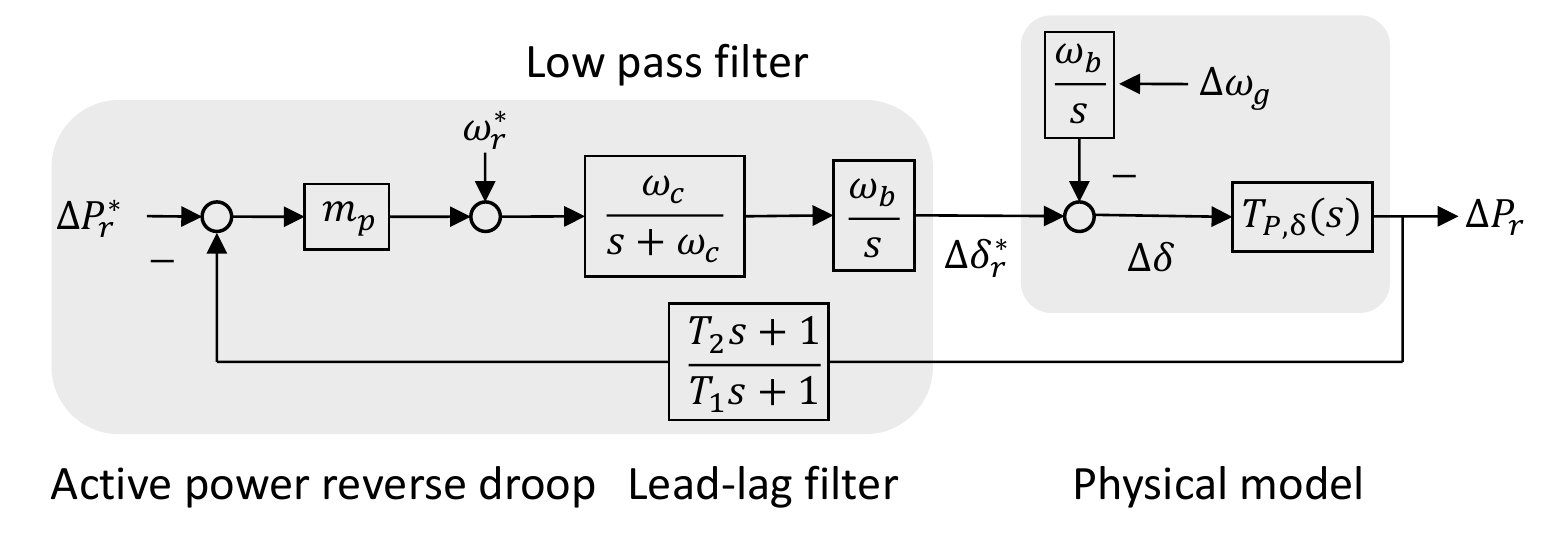}
	\caption{The MIGRATE EU project controller \cite{migrate3_2}.}\label{fig:grid-forming}
\end{figure}

The grid--supporting mode can be applied to a grid-- forming converter as presented in the MIGRATE EU project and depicted in Fig.~\ref{fig:grid-forming} \cite{migrate3_2}. Note that the equivalent model considers only the active power control loop since in the grid--forming converter there is a separation of the control channels (this is not true for the grid--following converter).
Parameter $m_p$ is defined as the inverse of the droop, $\omega_c$ is the cut--off frequency of the low pass filter added to increase the control performance by avoiding fast frequency variations. $T_1$ and $T_2$ are the time constants of the feedback function on the power measurements. Such a function adds a lead-lag action which improve the stability of the control \cite{migrate3_2}. 
As in \cite{migrate3_2}, from Fig.~\ref{fig:grid-forming}, the system transfer function can be computed as follows:
\begin{align}
    & \Delta P_r = G_{P_r,P_r^*} \Delta P^{*}_r + G_{P_r,\omega_r^*} \Delta \omega_r^* +G_{P_r,\omega_g} \Delta \omega_g ,\label{eq:gridfollowingG}\\
    & G_{P_r,P_r^*}=\frac{\Delta P_r}{\Delta P^{*}_r} = \frac{m_p G_1 T_{P,\delta }}{1+m_p G_1  T_{P,\delta } G_2} \nonumber,\\
    & G_{P_r,\omega_r^*}=\frac{\Delta P_r}{\Delta \omega_r^*} = \frac{G_1 T_{P,\delta }}{1+ G_1  T_{P,\delta } m_p G_2} \nonumber,\\
    & G_{P_r,\omega_g}=\frac{\Delta P_r}{\Delta \omega_g} = -\frac{T_{P,\delta } \omega_b/s}{1+ G_1  T_{P,\delta } m_p G_2} \nonumber,\\
    & G_1(s) = \frac{\omega_b \omega_c }{s(s + \omega_c)}, \quad \quad  G_2(s) = \frac{T_2 s +1}{T_1 s +1}. \nonumber
\end{align}

\subsubsection{Grid--following converter}
It behaves as an ideal current source connected to the grid. The current phasor is synchornized with the AC voltage through the \ac{pll}. The phase tracking is realised when the rotating frame angle $\vartheta_{pll}$ set by the PI controller is such that the \emph{q}-axis voltage results to be null \cite{Wu2018pll_stability}.

The grid coupling can be represented in terms of phasors and in the rotating frame of the \ac{pll}, as follows:
\begin{align}      
    & E_g e^{j\delta-\vartheta_{pll}} = I_g Z_g e^{j \varphi + \vartheta_z} + V_g e^{j \vartheta_g-\vartheta_{pll}},
 \end{align}   
where $\dot{z}_g = Z_g e^{j \vartheta_z}$ is the transformer impedance (see appendix~\ref{appendix:A}, Fig.~\ref{fig:diagramma_e_vettori}-a). The \emph{q} component can be expressed as
 \begin{align} 
    & e_{gq}=E_g \sin{\left(\delta-\vartheta_{pll}\right)}, \label{eq:gridformingG}\\
    & e_{gq}= I_g Z_g \sin{\left( \varphi + \vartheta_z\right)} + V_g \sin{\left( \vartheta_g-\vartheta_{pll}\right)}. \label{eq:egq}
\end{align}
Equation \eqref{eq:egq} can be linearized as
 \begin{align} 
    &  \Delta e_{gq} \approx  V_g {\left( \Delta \vartheta_g- \Delta \vartheta_{pll}\right)}, \label{eq:lin_egq}
\end{align}
since $I_g Z_g \sin{\left( \varphi + \vartheta_z\right)}$ can be assumed as constant. Then considering what described in Fig.~\ref{fig:grid_forming_feeding_control_loop}-a and that 
$\Delta \vartheta_g = 1/s \omega_g$, $\Delta \vartheta_{pll} = 1/s \omega_{pll}$ and $ \Delta \omega_{pll} = PI_{pll}(s) \Delta e_{gq} $
it is can be verified that the \ac{pll} behaviour can be reproduced by the block diagram in Fig.~\ref{fig:grid-feeding}.
Details can be found in appendix~\ref{appendix:A}.
\begin{figure}
	\centering
	\includegraphics[width=0.9\columnwidth]{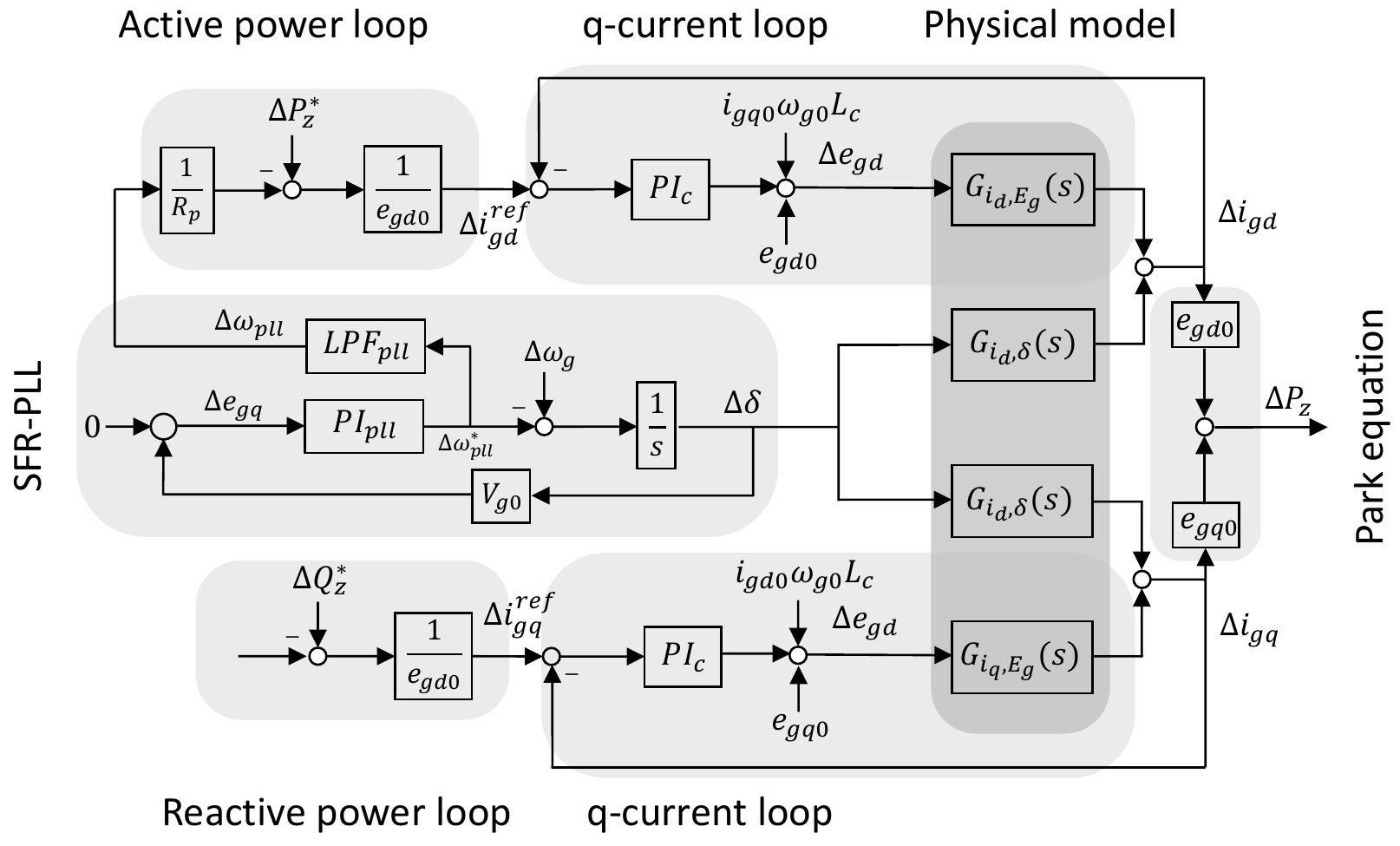}
	\caption{Grid--following equivalent linear model.}\label{fig:grid-feeding}
\end{figure}
In particular, the \ac{pll} behaviour can be represented with the following transfer functions:
\begin{align}
    \Delta \delta &=  \frac{1}{s + PI_{pll} V_{g0} }\Delta \omega_g = G_{\delta,\omega_g} \Delta \omega_g ,\\
    \Delta \omega_{pll} &= LPF_{pll}\frac{1/s \cdot PI_{pll} V_{g0}}{1+1/s \cdot PI_{pll} V_{g0}} \Delta \omega_g \nonumber\\
    &=G_{\omega_{pll},\omega_g} \Delta \omega_g.
\end{align}
where, $PI_x(s) = K_{p,x} + K_{i,x}/s$, and $LPF_{pll}(s)$ is the low pass filter transfer function. In this case we consider a second order filter with damping ration $\zeta$ and natural frequency or cut--off frequency $\omega_{lpf}$.

The control scheme of a grid--followiing converter does not exhibit the separation of active and reactive power control loops, therefore both channels should be considered in the model.
The current loop transfer function can be obtained, considering the block diagram in Fig.~\ref{fig:grid-feeding}, as:
\begin{align}
    & \Delta i_{gd} = \frac{PI_c G_{i_{d},E_g}}{1+PI_c G_{i_{d},E_g}} \Delta i_{gd}^{ref} +  \frac{G_{i_{d},\delta}}{1+PI_c G_{i_{d},E_g}}  \Delta \delta, \label{eq:curr_var_feeding}\\ 
    & \Delta i_{gq} = \frac{PI_c G_{i_{q},E_g}}{1+PI_c G_{i_{q},E_g}} \Delta i_{gq}^{ref} +  \frac{G_{i_{q},\delta}}{1+PI_c G_{i_{q},E_g}}   \Delta \delta.
\end{align} 
Note that since this is a variation model, the feed--forward control has been neglected. The current reference are:
\begin{align}
    & \Delta i_{gd}^{ref} =  \frac{1}{e_{gd0}} \Delta P_z^{*} - \frac{1}{R_p e_{gd0}} G_{\omega_{pll},\omega_g} \Delta \omega_g, \nonumber \\
    & \Delta i_{gq}^{ref} =  \frac{1}{e_{gd0}} \Delta Q_z^{*}.\label{eq:curr_ref_feeding}
\end{align} 
Thus, combining equations \eqref{eq:curr_var_feeding}--\eqref{eq:curr_ref_feeding} we derive: 
\begin{align}
    & \Delta i_{gd} = \frac{PI_c G_{i_{d},E_g}}{1+PI_c G_{i_{d},E_g}} \frac{1}{e_{gd0}} \Delta P_z^{*}  + \nonumber \\  & + \left( \frac{G_{i_{d},\delta} G_{ \delta,\omega_g }}{1+PI_c G_{i_{d},E_g}}   - \frac{PI_c G_{i_{d},E_g}}{1+PI_c G_{i_{d},E_g}}  \frac{ G_{\omega_{pll},\omega_g} }{R_p e_{gd0}} \right) \Delta \omega_g,
\end{align}    
\begin{align}   
    & \Delta i_{gq} = \frac{PI_c G_{i_{q},E_g}}{1-PI_c G_{i_{q},E_g}} \frac{1}{e_{gd0}} \Delta Q_z^{*}  +  \frac{ G_{i_{q},\delta}  G_{ \delta,\omega_g }}{1+PI_c G_{i_{q},E_g}} \Delta \omega_g.
\end{align} 
Finally, the active power variation can be computed as $\Delta P_z = \Delta i_{gd} e_{gd0} + \Delta i_{gq} e_{gq0} $, therefore,
\begin{align}
    & \Delta P_z = G_{P_z,P_z^{*}}  \Delta P_z^{*}  + G_{Q_z,Q_z^{*}} \Delta Q_z^{*} +   G_{P_z,\omega_g} \Delta \omega_g , \\
    & G_{P_z,P_z^{*}}  = \frac{PI_c G_{i_{d},E_g}}{1+PI_c G_{i_{d},E_g}}\nonumber, \\
    & G_{Q_z,Q_z^{*}} =\frac{PI_c G_{i_{q},E_g}}{1-PI_c G_{i_{q},E_g}} \frac{ e_{gq0}}{e_{gd0}} \nonumber ,\\
    & G_{P_z,\omega_g}  =  \nonumber \\ & \frac{ \left(G_{i_{d},\delta}e_{gd0} + G_{i_{q},\delta} e_{gq0}\right) G_{ \delta,\omega_g } }{1+PI_c G_{i_{d},E_g}} - \frac{PI_c G_{i_{d},E_g}}{1+PI_c G_{i_{d},E_g}}  \frac{ G_{\omega_{pll},\omega_g} }{R_p }. \nonumber
\end{align}  

\subsection{Hydroelectric power plant}

The hydroelectric unit can be represented by the following transfer function which considers the governor, with transient droop compensation modelled by the first two transfer functions, and the hydraulic turbine  \cite{kundur1994power,ieee1992hydraulic}:
\begin{align}
    & G_{j}(s) = \frac{\Delta P_{j}(s)}{\Delta P^*_{j}(s) + \Delta \omega_g(s)/R_j} \label{eq:hydroTF}\\ 
    & G_{j}(s) = \frac{K_{g,j}}{T_{g,j} s + 1}\cdot \frac{T_{r,j} s + 1}{\frac{R_{t,j}}{R_j}T_{r,j} s + 1}\cdot \alpha_{3,j}  \frac{- \alpha_{2,j} T_{w,j} s + 1}{ \alpha_{1,j} \frac{T_{w,j}}{2} s + 1} \nonumber 
  \end{align}
where $T_{g,h}$ is the governor time constant; $R_t$ and $T_r$ are the temporary droop and the reset time, respectively; 
$T_w$ is the water time constant, also said water starting time.

Coefficients $\alpha_1, \alpha_2, \alpha_3$ are related to the nonlinearity of the physical mode. In the literature there are several definitions mainly related on the hypothesis made to linearize the model, \emph{e.g.} in \cite{ieee1992hydraulic} it is said that the first two are proportional to the opening grade of the valve and therefore proportional to the initial active power while the third one is set equal to  one, thus, $\alpha_1=\alpha_2=k P_0, \alpha_3=1$, with $k=1/g_{max}-g_{min}$, namely, the maximum and the minimum opening of the control valve.
 
\subsection{Thermoelectric power plant}  

The thermoelectric power plant considered in this paper consists of a steam turbine of common use \cite{kundur1994power}. The system is composed by high pressure (HP) section, reheater and low pressure (LP) sections. The system controller acts on a control valve which modulates the steam in the flow in the HP section, this is modelled by the time constant $T_{ch}$. The flow which goes out of the HP section pass through an intercept valve which is ahead of the reheater. The time constant associated to the reheater is $T_{rh}$.  The power generated by the turbine is the sum of the power in each sections $F_{hp}$ and $F_{lp}$ indicate the per unit fraction of the power generated by the turbine sections when the valve is fully open and the turbine is generating the nominal power, therefore $F_{hp}+F_{lp}=1$ \cite{kundur1994power}. $T_{g,t}$ is the governor time constant. The complete transfer function is,
\begin{equation}
    G_{k}(s) = \frac{\Delta P_{k}(s)}{\Delta P^*_{k}(s) + \Delta \omega_g(s)/R_k} \label{eq:thermTF},
\end{equation}
\begin{equation}
G_{k}(s) = k_{g,k} \frac{T_{g2,k} s + 1}{T_{g1,k} s + 1} \cdot \beta_3 \frac{F_{hp,k} \beta_1 T_{rh,k} s +1}{\left(\beta_2 T_{rh,k} s + 1\right)\left(T_{ch,k} s + 1\right)} . 
\end{equation}
As for the hydro units, $\beta_1, \beta_2,\beta_3$ are parameters related to the linearization hypothesis. In this paper, we assume the definition reported in \cite{kundur1994power}, which is $\beta_1=\beta_2=1, \beta_3 = k P_0$.
  
\subsection{Control area power system} \label{sec:areaSS}
The control area power system can be approximated with a single bus--bar element whose response is approximated by a first order transfer function as shown in Fig.~\ref{fig:areamodel}, in which $H_i$ is the summation of all the inertia constant of the machine in the \emph{i-th} area, and $D_i$ is the damping coefficient. The interaction between the areas, can be modelled thought the tie--lines synchronising coefficients $T_{i,j}$, computed form the load flow results, \emph{e.g.} for a generic tie--line from area 1 to 2 \cite{bevrani2014robust}: 
$T_{12} = (|V_1||V_2|)/X_{12}\cos{(\delta_1^0-\delta_2^0)}$,
where  $X_{12}$ is the line reactance, $V_1,V_2$ are the voltages amplitude and $\delta_1^0,\delta_2^0$ are the initial angles.

\section{Parameters identification} \label{sec:Parameters_identification}
The parameters of the models presented in Section~\ref{sec:Models} are inferred by performing a system identification fed by measurements supposed to be available to the system operator from a distributed sensing infrastructure composed, for instance, by \ac{pmu}. 
Additionally, the identification problem requires the knowledge of the system initial condition also supposed to be available from the load flow calculation or a state estimation process.
More specifically, the measurements required for the algorithm consists of powers and frequency recorded at the occurrence of some events in the grid \emph{e.g.} load connections or disconnections,  variations of generator set-points.

Parameters of hydro and thermal units are generally unknown since they derive from the simplification and the linearization of complex physical systems. However, in the existing literature, some indications regarding the range of values for the parameters of equations~\eqref{eq:hydroTF} and \eqref{eq:thermTF} can be found in \cite{kundur1994power,saccomanno2003electric,ieee1992hydraulic}. These intervals can be used to constrain the identification problem.

Differently, the parameters required to determine the converters transfer functions can be derived using the values of gains and time constants, precisely defined in the devices design, and load flow results. In this work, we first assume the availability of such converters design parameters. However, in some cases, the system operator, which is the end--user of the proposed model, cannot have access to the values of these parameters. Therefore, an identification procedure is proposed both for hydro and thermal units and for converter--based units.

The identification is realised in two steps: first step: the unit transfer functions are identified by considering the single unit alone; second step: the grid transfer function is obtained by considering all the pre--identified models together.
\subsection{Step 1: Single unit function identification} \label{sec:subLSO1}
\subsubsection{Hydro and thermal units}
The machine parameters identification is performed by solving the following \ac{lso} problem for each $j$-th hydro unit and $k$-th thermal unit. The solution space is box-constrained by literature--based range values.
\begin{align}
    & \min_{\vec{x}_{h,j,(t,k)}}    \sum_{t_s} (P_{h,j,(t,k)}(t_s) - \hat{P}_{h,j,(t,k)}(t_s))^2 \label{eq:LSO_machines1} \\
    & \vec{x}_{h,j} = \begin{bmatrix} T_{g,j} & T_{r,j} & R_{t,j} \end{bmatrix}\\
    & \vec{x}_{t,k} = \begin{bmatrix} T_{g1,k} & T_{g2,k} & T_{rh,k} & T_{ch,k} & T_{hp,k} \end{bmatrix} \\
    & \vec{LB}_{h,j,(t,k)} \leq \vec{x}_{h,j,(t,k)} \leq \vec{UB}_{h,j,(t,k)} \\
    & P_{h,j,(t,k)}(t_s) = \text{lsim}\left(G_{h,j (t,k)}(s) , \hat{u}_{h,j,(t,k)}(t_s)\right) \label{eq:LSO_machines2}
\end{align} 
The optimisation variables are $\vec{x}_{h,j}$ for the hydro case and $\vec{x}_{t,k}$ for the thermal one. $\vec{UB}_{h,j,(t,k)}$ and $\vec{LB}_{h,j,(t,k)}$ are the upper and lower bound which delimited the space of $\vec{x}$. $t_s$ is the sampling time.
Moreover, the function \textit{lsim(transfer function, input)} indicates the results of the simulation of the the given transfer function $G_{h,j (t,k)}(s)$ parametrized with the $\vec{x}_{h,j,(t,k)}$, for a given input. Specifically, in this context the input corresponds to the regulating signal provided to the units, $\Delta P^*_{h,j,(t,k)}(t_s) + \Delta\omega_g(t_s)/R_{j,(k)}=\hat{u}_{h,j,(t,k)}(t)$ and the measured active power corresponding to the same input signal, $\hat{P}_{h,j,(t,k)}(t)$. 
Problem \eqref{eq:LSO_machines1}--\eqref{eq:LSO_machines2} is nonlinear and non-convex and, in order to detect the best results, we adopted a brute-force approach where several LSO have been solved starting from different initial conditions, $\vec{x}^0_{h,j,(t,k)}$, randomly defined within the solution space.
Thus, the best solution is the one that ensures the best fitting between the real $\hat{P}_{h,j,(t,k)}(t_s)$ and the measured power $P_{h,j,(t,k)}(t_s)$. 

\subsubsection{Converter-based units}\label{ssec:coverter-based_units_identification}
For converter--based units, the parameter identification is performed by solving the following \ac{lso} problem for each $l$-th grid--following and $m$-th grid--forming converter--based unit.
\begin{align}
    & \min_{\vec{x}_{r,l,(z,m)}}  \sum_{t_s} (P_{r,l,(z,m)}(t_s) - \hat{P}_{r,l,(z,m)}(t_s))^2 \label{eq:LSO_conv1}\\ 
    & \vec{x}_{r,l} = \begin{bmatrix} \omega_{c,r,l} & T_{1,r,l} & T_{2,r,l} \end{bmatrix}\\
    & \vec{x}_{z,m} = \begin{bmatrix} k_{i,pll} & k_{p,pll} & k_{i,p} & k_{i,i} & \omega_{lpf} \end{bmatrix}\\
    & \vec{LB}_{r,l,(z,m)} \leq \vec{x}_{r,l,(z,m)} \leq \vec{UB}_{r,l,(z,m)} \\
    & P_{r,l,(z,m)}(t_s) = \text{lsim}\left(G_{r,l,(z,m)}(s) , \hat{u}_{r,l,(z,m)}(t_s)\right) \label{eq:LSO_conv2}
\end{align}
\noindent where $ G_{r,l}(s)$, $ G_{z,m}(s)$ and $\hat{u}_{r,l}(t_s)$, $\hat{u}_{z,m}(t_s)$, are the transfer functions and the input vectors of the multi-input models \eqref{eq:gridfollowingG} and \eqref{eq:gridformingG}, defined as follows:
\begin{align}
    & G_{r,l}(s) = \begin{bmatrix} G_{P_{r,l},P_{r,l}^*} & G_{P_{r,l},\omega_{r,l}^*} &  G_{P_{r,l},\omega_g} \end{bmatrix}\\
    & G_{z,m}(s) = \begin{bmatrix}  G_{P_{z,m},P_{z,m}^{*}} & G_{Q_{z,m},Q_{z,m}^{*}} &   G_{P_{z,m},\omega_g}  \end{bmatrix}\\
    & \hat{u}_{r,l}(t_s) = \begin{bmatrix} \Delta P^{*}_{r,l}(t_s) & \Delta \omega_{r,l}^*(t_s) &  \Delta \omega_g (t_s) \end{bmatrix}^T\\
    & \hat{u}_{z,m}(t_s) = \begin{bmatrix}  \Delta P_{z,m}^{*}(t_s) & \Delta Q_{z,m}^{*}(t_s) &   \Delta \omega_g (t_s) \end{bmatrix}^T
\end{align} 
The optimization variables are $\vec{x}_{r,l}$, for grid--following converters and $\vec{x}_{z,m}$, for grid--forming converters. Note that these vectors collect only subsets of the parameters required by the two transfer functions. Indeed, they are the only ones that might be unknown by the system operator since the other parameters, $L_c$ and $R_c$, should be known in order to solve the load flow. Initial powers, currents and angles at the coupling point are provided by a load flow solution; and active and reactive power set points are assumed to be know by the system operator.

\subsection{Step 2: Grid transfer function identification}
The grid transfer function is identified by means of the following \ac{lso} problem:
\begin{align}
    & \min_{\vec{x}_{g}}    \sum_t^N \left(\omega_{g,i}(t) - \hat{\omega}_{g,i}(t)\right)^2 \label{eq:LSO_grid1} \\
    & \vec{x}_g = \begin{bmatrix} H_{i} & D_{i}  \end{bmatrix} \\
    & \vec{LB}_{g} \leq \vec{x}_{g} \leq \vec{UB}_{g} \\
    & \omega_{g,i}(t) = \text{lsim}\left(SYS(s), \hat{u}_{g,i}(t)\right).\label{eq:LSO_grid2}
\end{align} 
where $\vec{x}_{g}$ is a vector composed by the inertia constants and the damping coefficients of all the areas; $\hat{\omega}_{g,i}(t)$ is the frequency of the centre of inertia provided by the real--time measurements; $SYS(s)$ is the complete multi--area system transfer function; $\vec{LB}_{g}$ and $\vec{UB}_{g}$ are the upper and lower bound that define the solution space.
Specifically, for the inertia the limits are imposed by considering a variation of $\pm 20\%$ from the theoretical value. For the damping coefficients, the limits are imposed by considering the type of load \emph{i.e.}  commercial,  industrial and residential \cite{kundur1994power}. 
Let us consider the following load model and its linearization in respect to $\Delta \omega_g$,
\begin{align}
    & P_l = P_{l0}\left( \frac{V_g}{V_{n}}\right)^{k_{pv}} \left( 1+k_{pf} \omega_g \right) \\
    & \Delta P_l = P_{l0}\left( \frac{V_{g0}}{V_{n}}\right)^{k_{pv}} k_{pf} \Delta \omega_{g} = K \Delta \omega_{g} ,
\end{align} 
it can be demonstrated that the summation of all the terms $K$ for all the load in the area is equal to $D_i$. Therefore, by knowing the range of variation for $k_{pv} \in [k_{pv}^{min},k_{pv}^{max}]$ and $k_{pf} \in [k_{pf}^{min},k_{pf}^{max}]$ \cite{kundur1994power} and the load flow results for $V_{g0}$, it is possible to define a range  $D_i \in [D_i^{min} D_i^{max}]$.

\section{Simulation setup} \label{sec:real_time_simulator}
\begin{figure}[t]
	\centering
	\includegraphics[width=1\columnwidth]{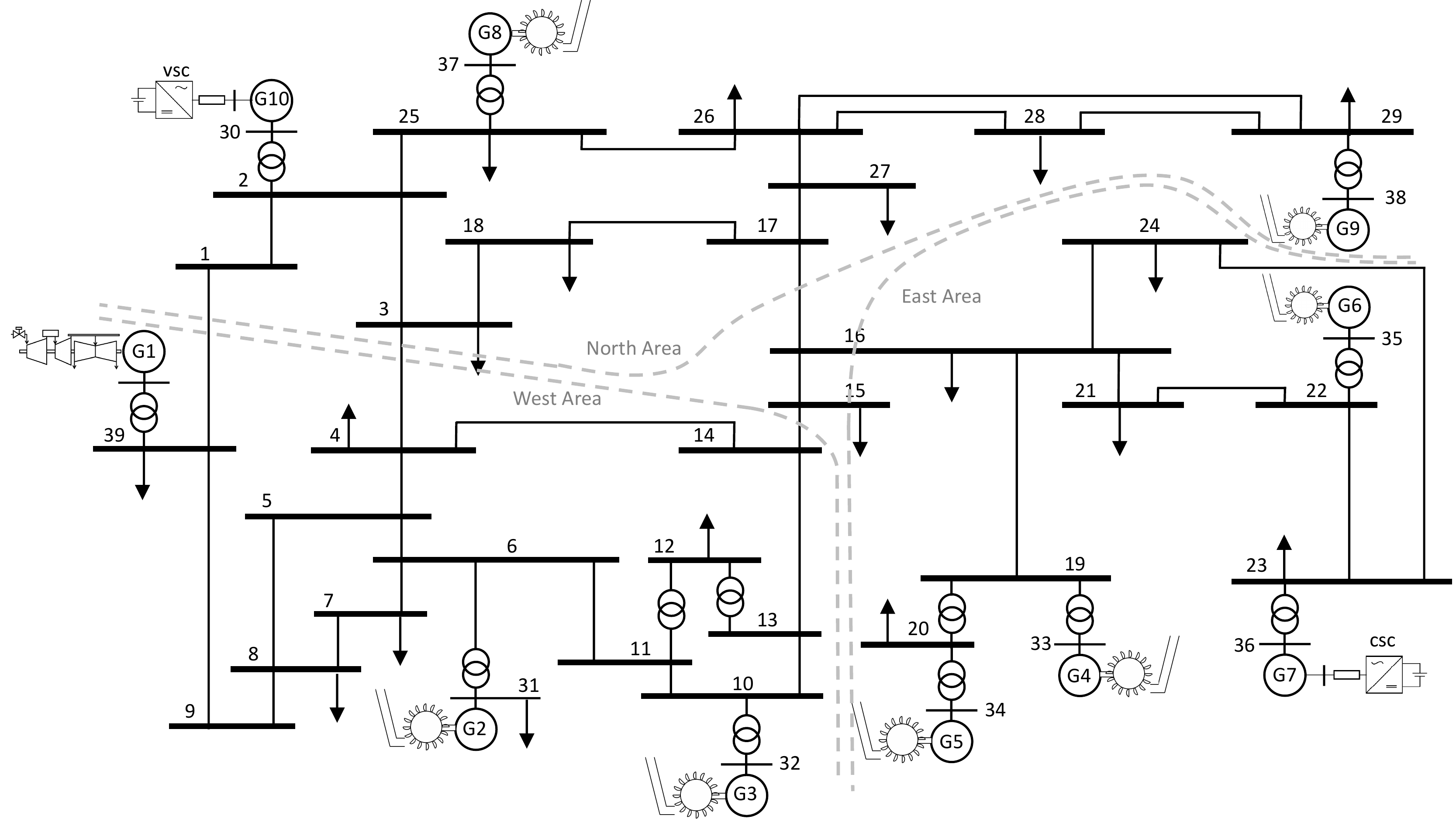}
	\caption{Modified IEEE 39 bus system \cite{ramos2015benchmark}.}\label{fig:IEEE39bus}
\end{figure}

The equivalent models and the identification procedure, introduced in the previous sections, are validated by using a  real–-time simulation setup which implements a  modified version of the IEEE 39 bus benchmark system, Fig.~\ref{fig:IEEE39bus} \cite{ramos2015benchmark}. Details on the parameters of lines, loads, machines and load flow can be found in \cite{ramos2015benchmark,dervivskadic2018under,zuo2018dispatch} and the grid model is available in open source at \cite{githubDESLEPFL}. The main variations form the benchmark system are listed here below:
\begin{itemize}
    \item synchronous machine nominal power is \SI{1000}{\mega\va} except for generator 1 (G1) which is rated \SI{3000}{\mega\va};
    \item in the study case configuration (Fig.~\ref{fig:IEEE39bus}) G1 is steam power plant, G7 and G10 are a grid--following and grid--forming converter--based units, respectively. All the other generators are driven by hydro units; 
    \item the base case configuration, considers G1 steam power plant and all the others hydro units;
    \item nominal frequency is \SI{50}{\hertz}, therefore the impedances and reactances are recomputed from the original \SI{60}{\hertz} system \cite{ramos2015benchmark}.
\end{itemize}
\begin{figure}[t]
	\centering
	\includegraphics[width=0.8\columnwidth]{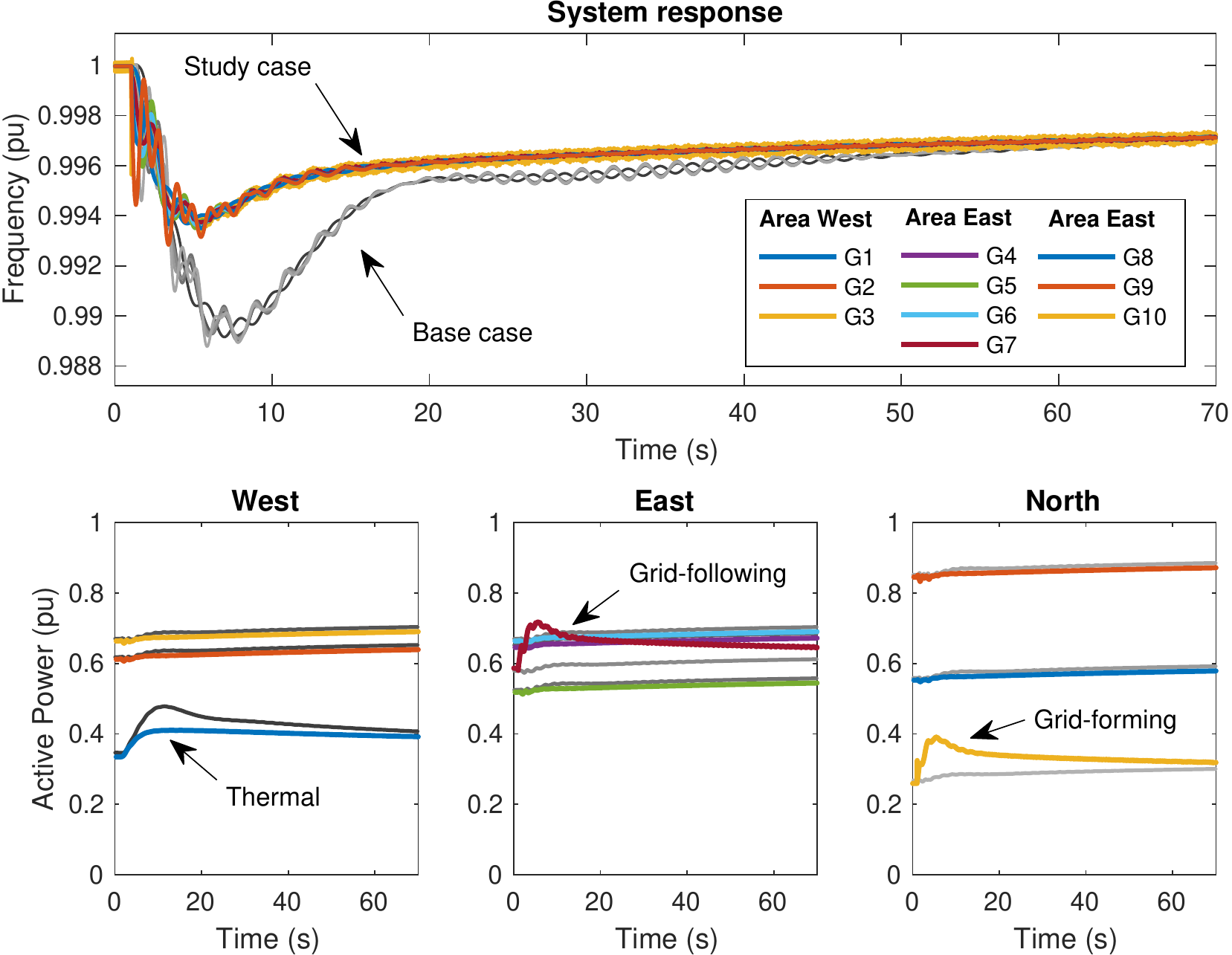}
	\caption{Real--time simulation of load change in bus 29, (from \SI{283.9}{\mega\watt} to \SI{783.5}{\mega \watt}).}\label{fig:confrontLoad23}
\end{figure}
The system is modelled in Matlab/Simulink and deployed on the Opal--RT platform\footnote{For this specific application the use of a real--time simulator is required in view of its high computational power. Indeed, the simulations performed on the Opal--RT platform are much faster than on a normal PC.}. Load flow calculations are performed with the Matpower toolbox.
Hydro units prime mover is realised with a nonlinear model under the assumption of non--elastic water column \cite{ieee1992hydraulic}. The reheated steam turbine model includes the speed governing system with proportional regulator, speed relay, and servomotor and the four-stage steam turbine each one modelled by a first-order transfer function \cite{kundur1994power}.
Synchronous generator and transformer models have been taken from the standard SimPowerSystems block-set. The loads are implemented as a static model with exponential dependency from voltage and proportional to frequency  \cite{kundur1994power}.
The converter units are modelled as presented in Fig.~\ref{fig:grid_forming_feeding_control_loop}. The DC part consists of a voltage source in series with a resistance and in parallel with a capacitance and  a controlled current source whose value is defined by the DC voltage controller of the DC bus. The main parameters of the converter units are reported in Appendix~\ref{appendix:B}.

Figure~\ref{fig:confrontLoad23} shows a transient in the base case (grey lines), and in the study case (coloured lines). Specifically, it consists of a step load change at bus 29, from the initial point of \SI{283.50}{\mega\watt} and \SI{26.90}{\mega\var}, the active power consumption is increased up to \SI{783.50}{\mega\watt}. 

\section{Results}\label{sec:results}

Four different grid transients, recorded with the study case configuration, have been considered:
\begin{itemize}
    \item[(a)] load increase at bus 29, from \SI{283.9}{\mega\watt} to \SI{783.5}{\mega \watt}, 
    \item[(b)] large load disconnection (\SI{628}{\mega\watt}, \SI{103}{\mega\var}) at bus 20,
    \item[(c)] grid--following converter set--point variation from $560$ \si{\mega\watt} to \SI{750}{\mega\watt}, 
    \item[(d)] grid--forming converter set--point variation from $250$ \si{\mega\watt} to \SI{100}{\mega\watt}.
\end{itemize}
To validate the approach, the accuracy of the single machines transfer functions identification (step 1) is quantified by comparing the power real--time measurements with the outputs of the identified models. Then, the full linear system is derived by applying step 2 of the identification procedure, and using the equivalent models of converters. The accuracy of this model is assessed by comparing its outputs --frequencies and powers-- with the simulated real--time measurements.
\begin{figure}[t]
	\centering%
	\includegraphics[width=0.9\columnwidth]{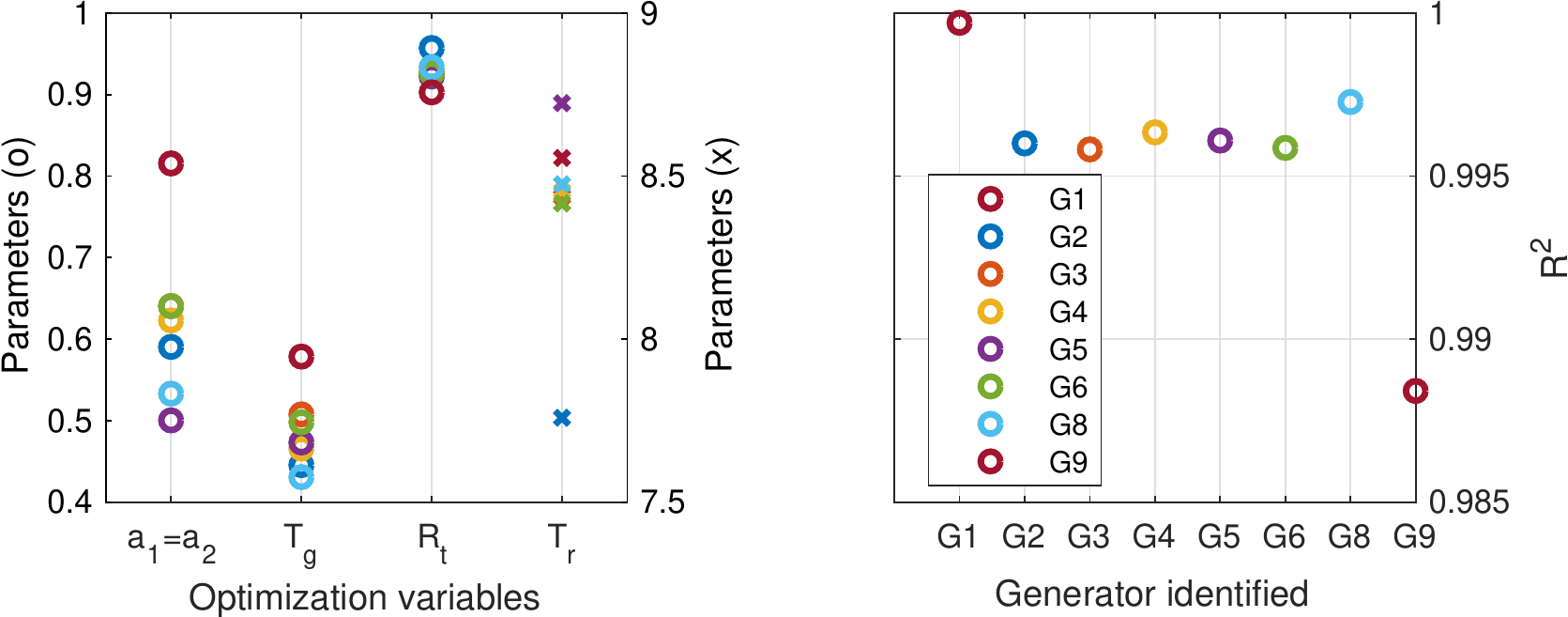}
	\caption{Machine parameters identification results for a load change at bus 29 (from \SI{283.50}{\mega\watt} to \SI{783.50}{\mega \watt}), in the study case configuration.}\label{fig:identificationLoad23}
\end{figure}

To evaluate the precision on the identification of the single machine parameters, we introduce the coefficient of determination $R^2$, defined as $ R^2 = 1-RSS/TSS$. $RSS = \sum_i{(y_i - \hat{y}_i)^2}$ is the so called residual sum of squares, $TSS = \sum_i{(y_i - \bar{y}_i)^2}$ is the total sum of squares, $\hat{y}_i$ are the observed data, $\bar{y}_i$ is the mean value of the observed data and $y_i$ are the data estimated by the model. 

Figure~\ref{fig:identificationLoad23}-left shows the result of the identification (step 1) performed for the hydro units only, in the case of transient (a). Figure~\ref{fig:identificationLoad23}-right shows the coefficient of determination $R^2$ for all identified generators, for the same transient event. We can observe that $R^2$ is always close to the unitary value, which indicates a high level of fitting. Similar results have been obtained with the other three transients.
\begin{figure*}
	\centering
	\subfloat[ ]{\includegraphics[width=0.5\columnwidth]{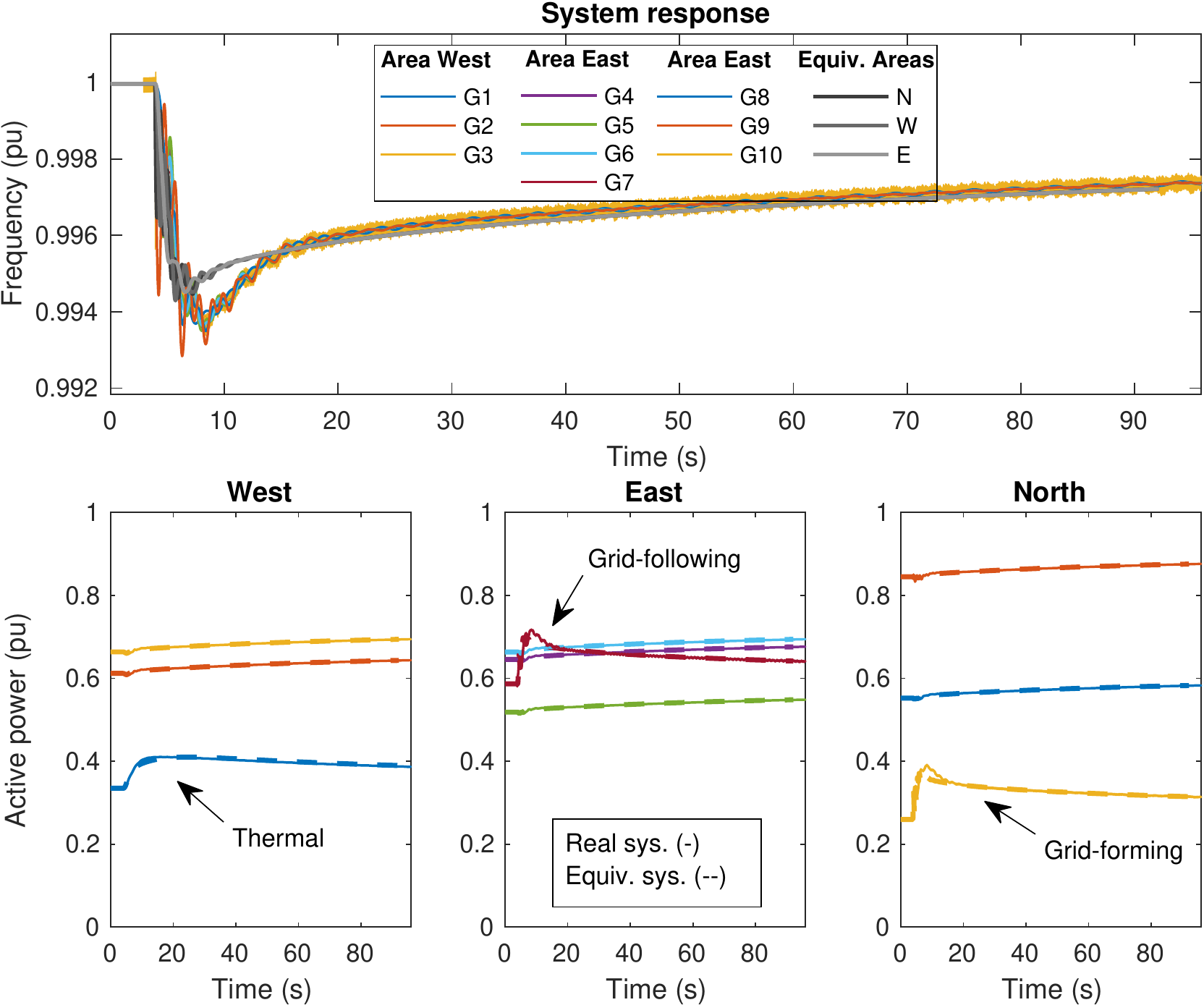}}   \hfil
	\subfloat[ ]{\includegraphics[width=0.5\columnwidth]{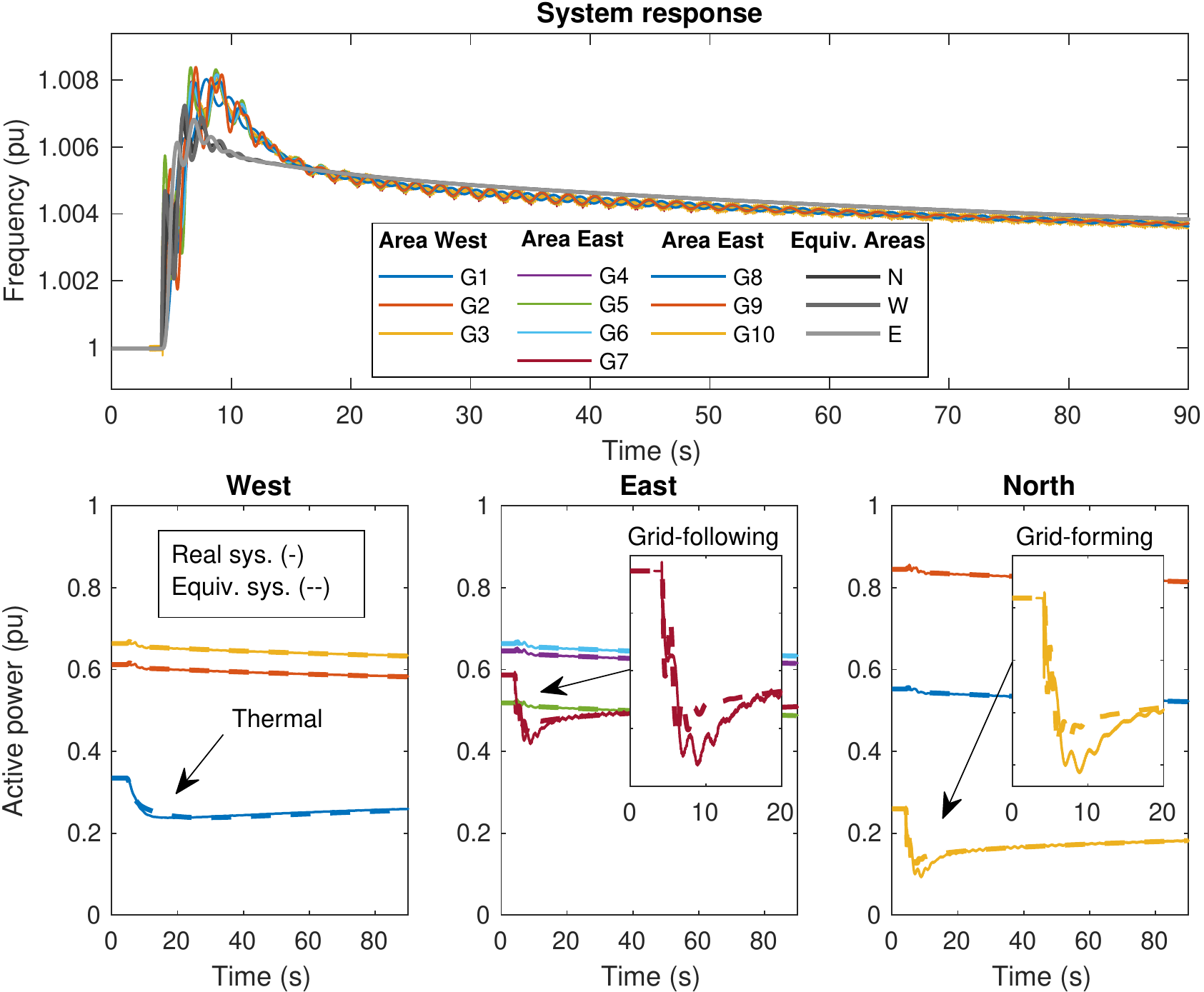}} \\
	\subfloat[ ]{\includegraphics[width=0.5\columnwidth]{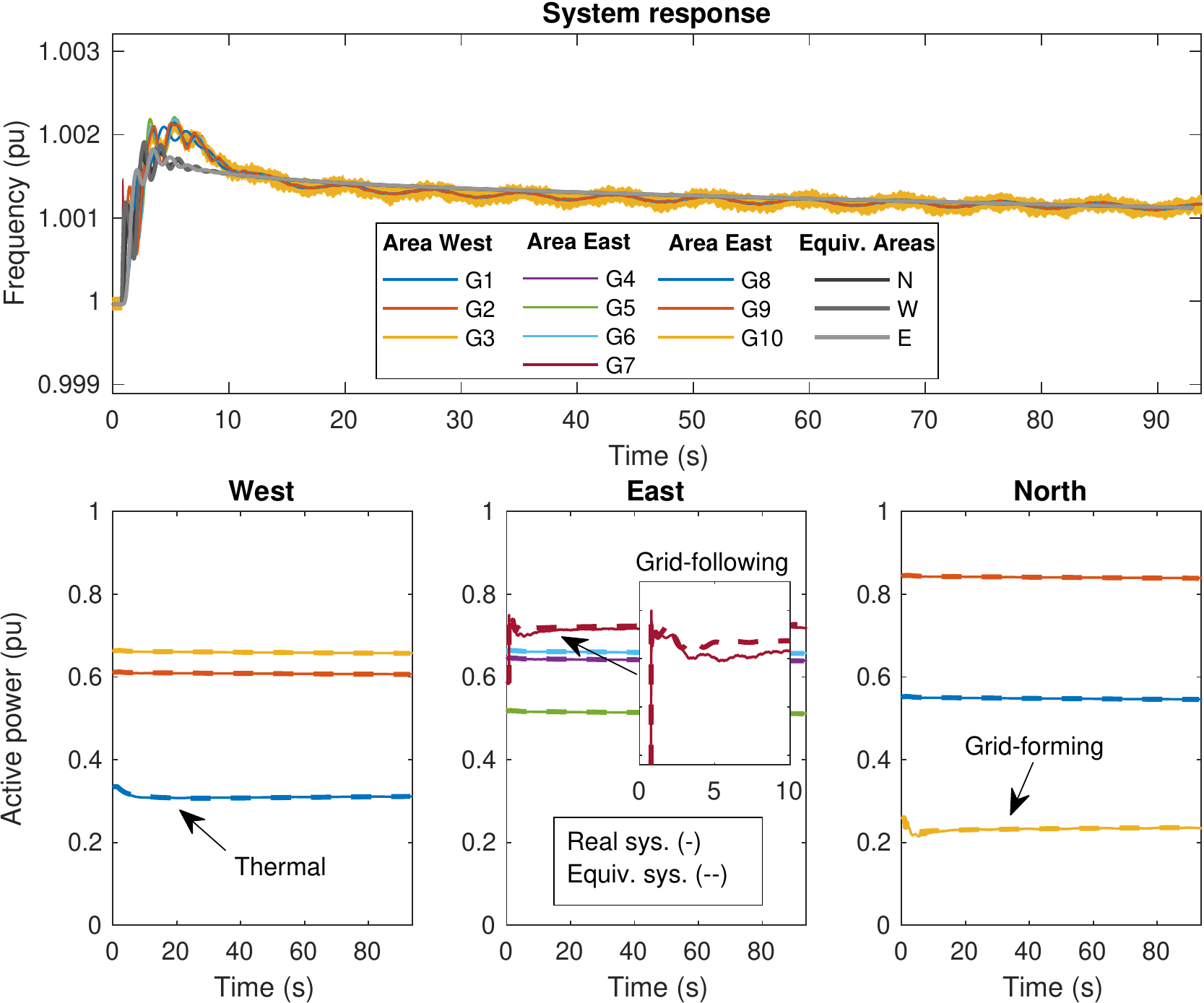}} \hfil
	\subfloat[ ]{\includegraphics[width=0.5\columnwidth]{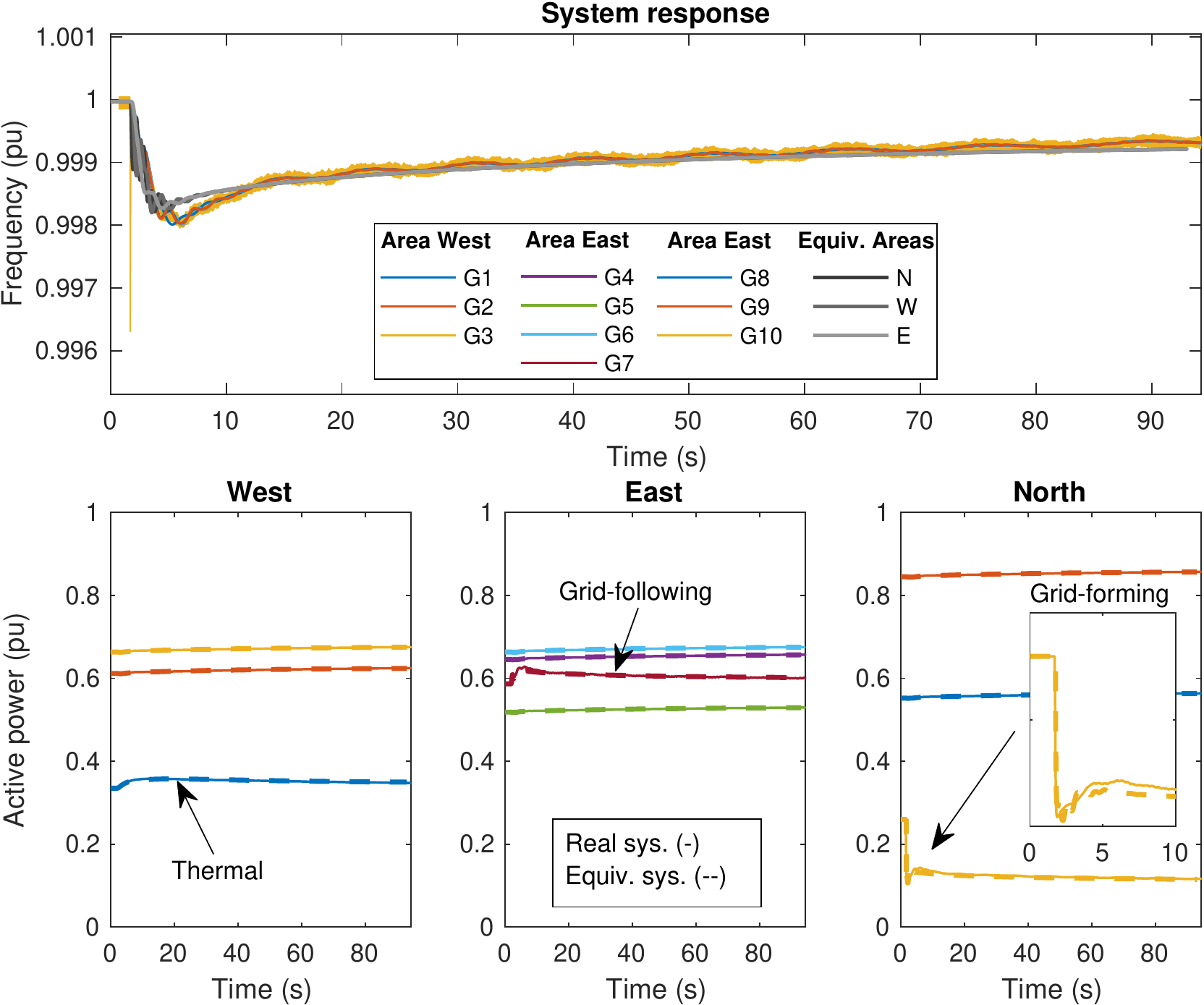}}
	\caption{Comparison of the outputs of the equivalent full linear model with the measurements of the real-time simulation: (a) load change in bus 29 (from \SI{283.9}{\mega\watt} to \SI{783.9}{\mega \watt}), (b) load 20 disconnection (\SI{628}{\mega\watt} \SI{103}{\mega\var}) (c) G7 set--point (from \SI{560}{\mega\watt} to \SI{750}{\mega\watt}) and (d) G10 set--point (from \SI{250}{\mega\watt} to \SI{100}{\mega\watt}).}\label{fig:identification}
\end{figure*}

The results of the full linear system, obtained for all the four transient events, (a) to (d), are depicted in Fig.~\ref{fig:identification}. 
For all of them, the top plot shows the frequencies measured from the detailed real-time simulation of machines and converters (coloured lines), compared with the frequencies returned from the three--area equivalent linear model (grey lines). The bottom plots reports the generated active power of the different resources divided per area; here, solid coloured lines are the real--time measurements, while grey dashed lines are related to the equivalent model. 
We can observe that the outputs of the equivalent linear model are accurate since the area frequencies are close to the measured ones. A high accuracy can be also detected on the power profiles. In particular, Fig.~\ref{fig:identification}-b, -c and -d highlight the power outputs of the converter models, deduced from theirs nominal parameters, which also shows a good fitting with the measurements.
\begin{figure}
	\centering
	\subfloat[ ]{\includegraphics[width=0.8\columnwidth]{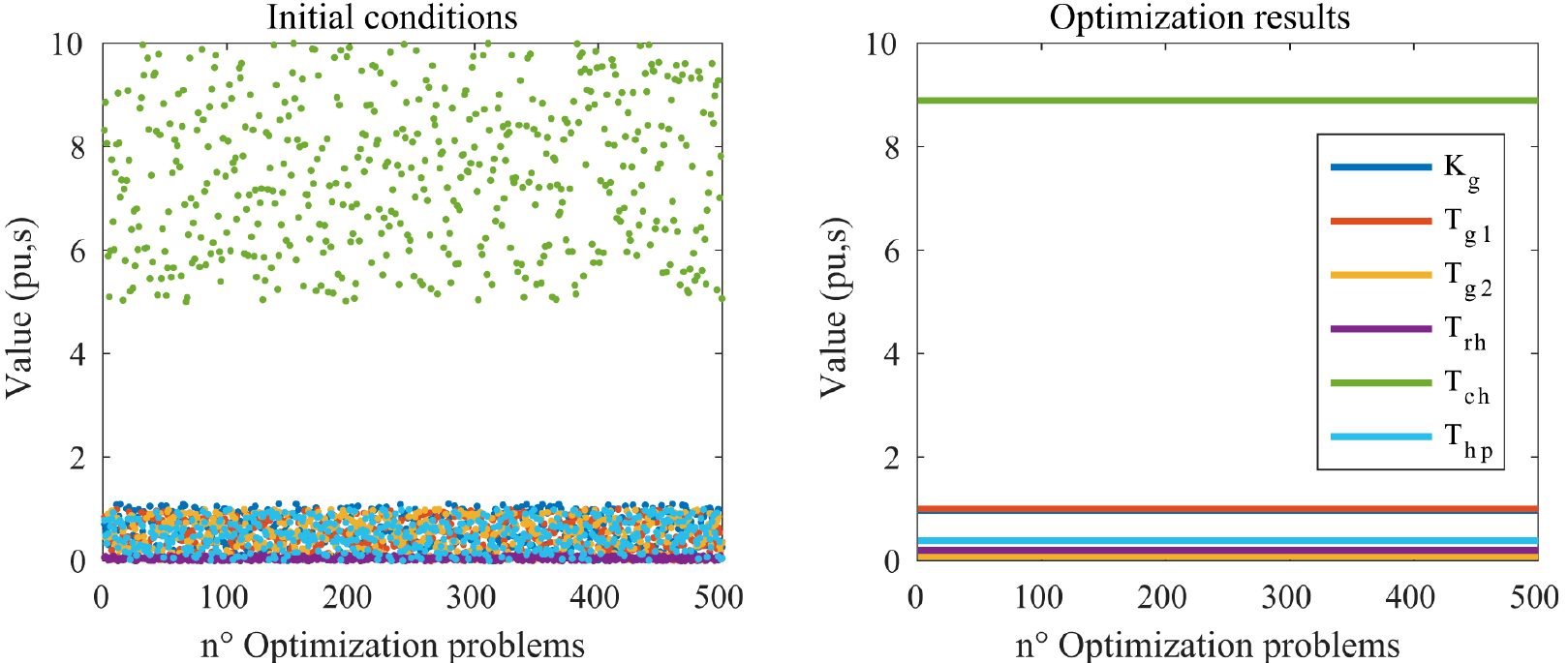}}\\
	\subfloat[ ]{\includegraphics[width=0.8\columnwidth]{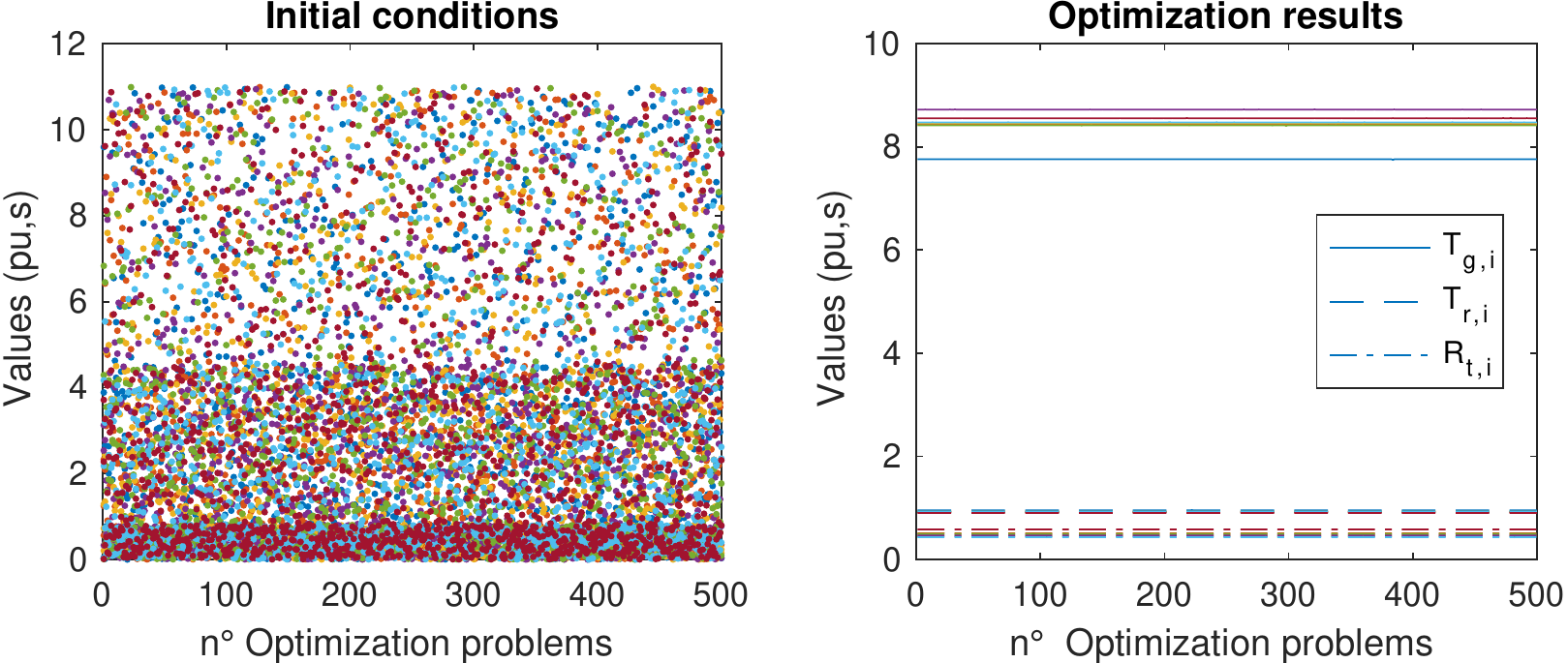}}\\
	\subfloat[ ]{\includegraphics[width=0.8\columnwidth]{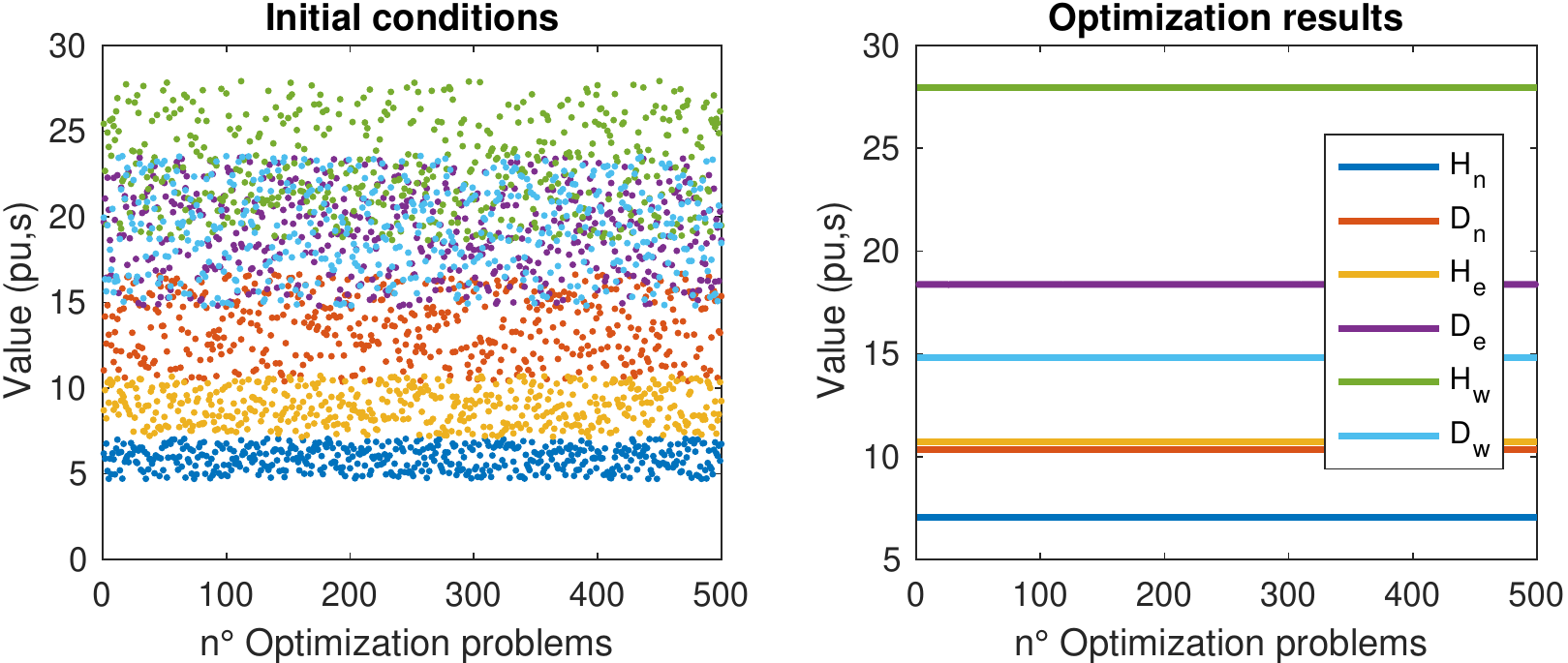}}
	\caption{Transfer functions identification, 500 iterations, case: load change in bus 29, (from \SI{283.50}{\mega\watt} to \SI{783.50}{\mega \watt}) (a) Thermoelectric, (b) Hydroelectric power plant and (c) power grid.}\label{fig:identification_500}
\end{figure}
\begin{figure}
	\centering
	\subfloat[ ]{\includegraphics[width=0.8\columnwidth]{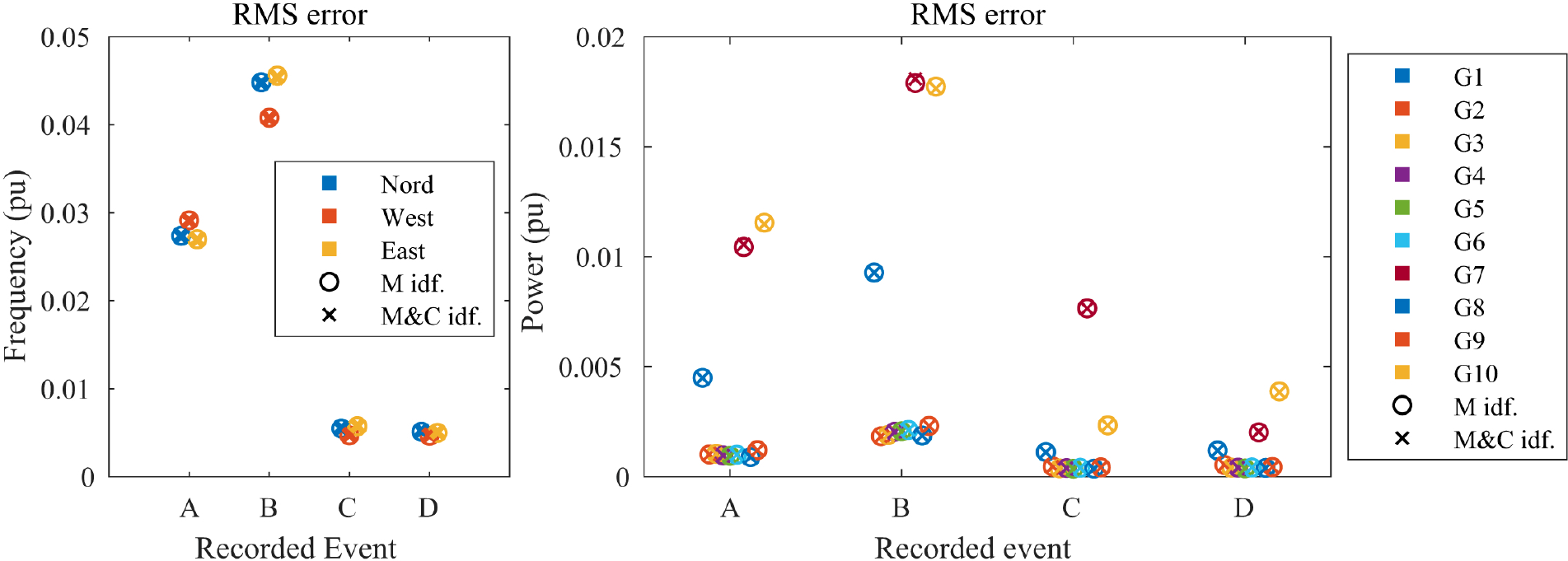}}\\
	\subfloat[ ]{\includegraphics[width=0.8\columnwidth]{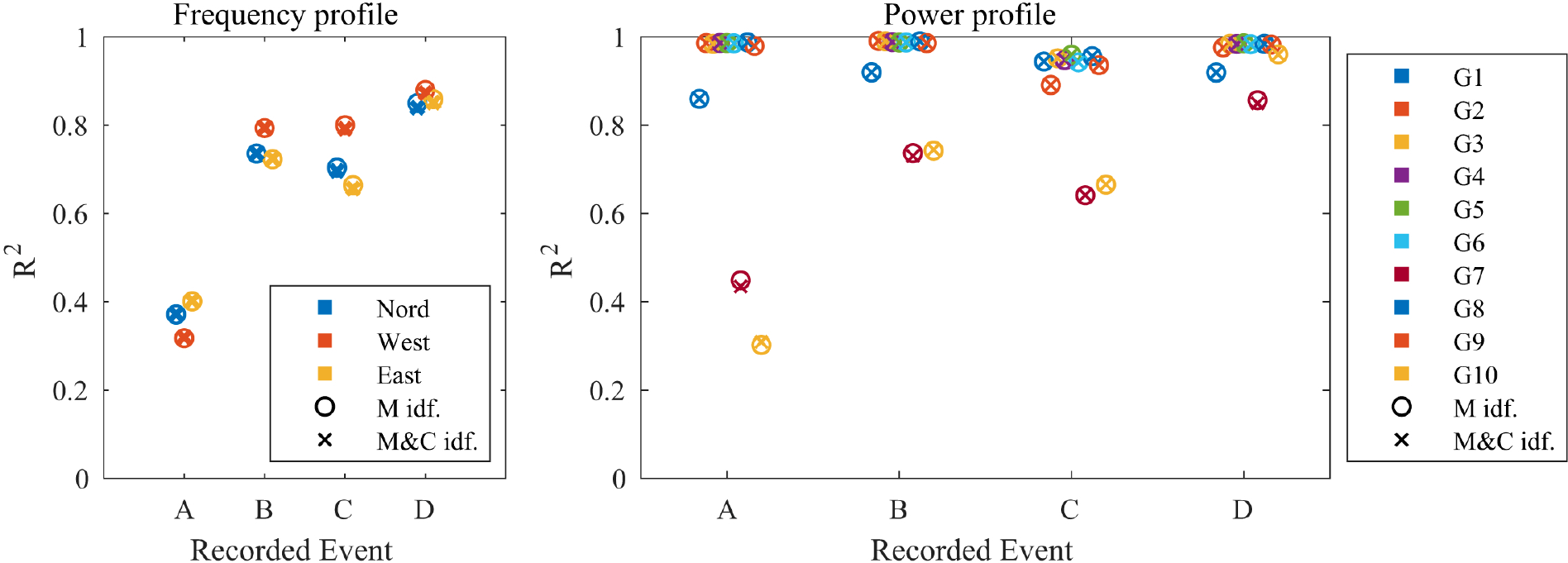}}
	\caption{Linear and non linear model results comparison (a) root men square error and (b) coefficient of determination.}\label{fig:conve_idf}
\end{figure}

The \ac{lso} problems, defined for the system identification, are, in principle, non-convex; therefore, the optimisation algorithm may not determine the global minimum of the objective function (entailing the identification of the most likelihood values of the parameters). Nevertheless, a non--convex problem may result to be convex within the solution space given by the constraints. Although we do not have a formal proof of this property, it seems to be the case for the presented problems. Indeed, Fig.~\ref{fig:identification_500} shows the results of 500 optimisation problems, specifically Fig.~\ref{fig:identification_500}--a presents on the left the initial points of each identification problem for the thermoelectric unit and on the right the optimal solution. As it can be seen for each initial point, selected randomly within the upper and lower bound, the algorithm converges to the same solution. The same happens for the hydroelectric units, Fig.~\ref{fig:identification_500}--b, and for the power grid identification, Fig.~\ref{fig:identification_500}--c.

Considering the practical implementation of the proposed framework, once the event is recorded, the whole model tuning is performed in almost \SI{175}{\second}. The first
\ac{lso} takes almost \SI{25}{\second} to perform the parameters identification of the 8 traditional units, while the second \ac{lso} computes the area control model in almost \SI{150}{\second}. Once the model is computed, each simulation of the linear model takes less than \SI{1}{\second} to be executed\footnote{The computation times discussed in this subsection has been measured on a MacBook Pro with \SI{2.50}{\giga\hertz} GHz Intel Core i7 dual-core processor and \SI{16}{\giga\byte} of RAM.}.Table~\ref{tab:GridIdentif} shows the numerical results of the grid transfer function identification. Most of the values, computed for the different simulations, are similar each other. 
Therefore, it could be reasonable not to evaluate at each identification the grid transfer function parameters. A dynamic identification can then be done by using the full procedure only one time, and then performing the model updating using only the identification of the machines’ transfer functions.

Based on these considerations, the model update for a large power system can be operated with a rate in the order of minutes. For example, the update can be carried out every time there is a change in the network operating conditions (\textit{e.g.} each hour for many real power systems) and every time a significant frequency variation is detected.

\begin{table}
		\centering
		\caption{Grid transfer function identification. Where $H_i$ is the inertia constant defined as $H_i =  J_i \omega_0^2/{S}_{base}$, and $D$ is the damping coefficient \cite{kundur1994power}.}
		\label{tab:GridIdentif}
		 \renewcommand{\arraystretch}{1.09}
		 \begin{tabular}{l l l l l}
		 \hline \hline
		 Variable   & Case (a)  & Case (b)  & Case (c) & Case (d) \\\hline 
	     $H_1$ [\si{\second}]     &  7.0560   &  7.0560   &  7.0560  &  7.0560\\
	     $D_1$ [\si{\pu}]    & 14.1662   & 10.3428   & 14.9381  & 10.3428\\\hline 
	     $H_2$  [\si{\second}]     & 10.7280   & 10.7280   & 10.7280  & 10.7280\\
	     $D_2$ [\si{\pu}]     &  17.1244  & 21.7686   & 16.1552  & 16.1856\\\hline
	     $H_3$ [\si{\second}]      & 27.9320   & 27.9320   & 27.9320  & 24.7406\\
	     $D_3$ [\si{\pu}]     & 14.8095   & 14.8095   & 15.9328  & 15.1430\\
		 \hline \hline
        \end{tabular}
\end{table}

A further scenario with unknown converters parameters, identified through the procedure described in subsection \ref{ssec:coverter-based_units_identification}, has been considered. Figure~\ref{fig:conve_idf} shows the \ac{rms} and the coefficient of determination $R^2$ both computed on the difference between frequencies and powers obtained with the identified linear model and with the reference nonlinear model. Results with the 'o' symbol refer to the case where the identification is performed only for traditional units; results with the 'x' symbol refer to the case where also converters parameters are identified. It can be noted that results obtained with and without the availability of the converter parameters are vary close each other. This proves the effectiveness of the approach also when the system operator has no information about the converter--based units parameters. 

\section{Conclusions} \label{sec:Conclusions}
The paper has proposed simplified linear models of grid--forming and grid--following converters, coupled with a least-square-optimisation-based method capable to identify the parameters of the proposed models. The objective is to obtain an equivalent linear model able to represent the power-frequency dynamical behaviour of a power system in which both traditional and converter-based units are involved in the control scheme. The approach has been validated thought comparisons with a full-replica model of the IEEE 39 bus system deployed on the Opal--RT simulation platform.
The proposed equivalent models have proved their capability to track the original system dynamics with high fidelity for all the considered scenarios.
Moreover, the paper shown the identifiability of the parameters characterising the proposed simplified linear models.
First, the consistency of the machines parameter estimates has been certified by observing coefficients of determinations close the unitary value. Moreover, the non convex identification problem has been numerically verified to be locally convex by limiting the identified parameters with bounds inspired by the technical literature. 

Future studies will focus on the experimental validation of the approach using real-field measurements. Moreover, the application of the equivalent linear model for extended system frequency stability studies will be studied.

\appendix
 \section{Phasor transient analysis} \label{appendix:A}
\begin{figure}
\centering
\subfloat[ ]{\includegraphics[width=0.5\columnwidth]{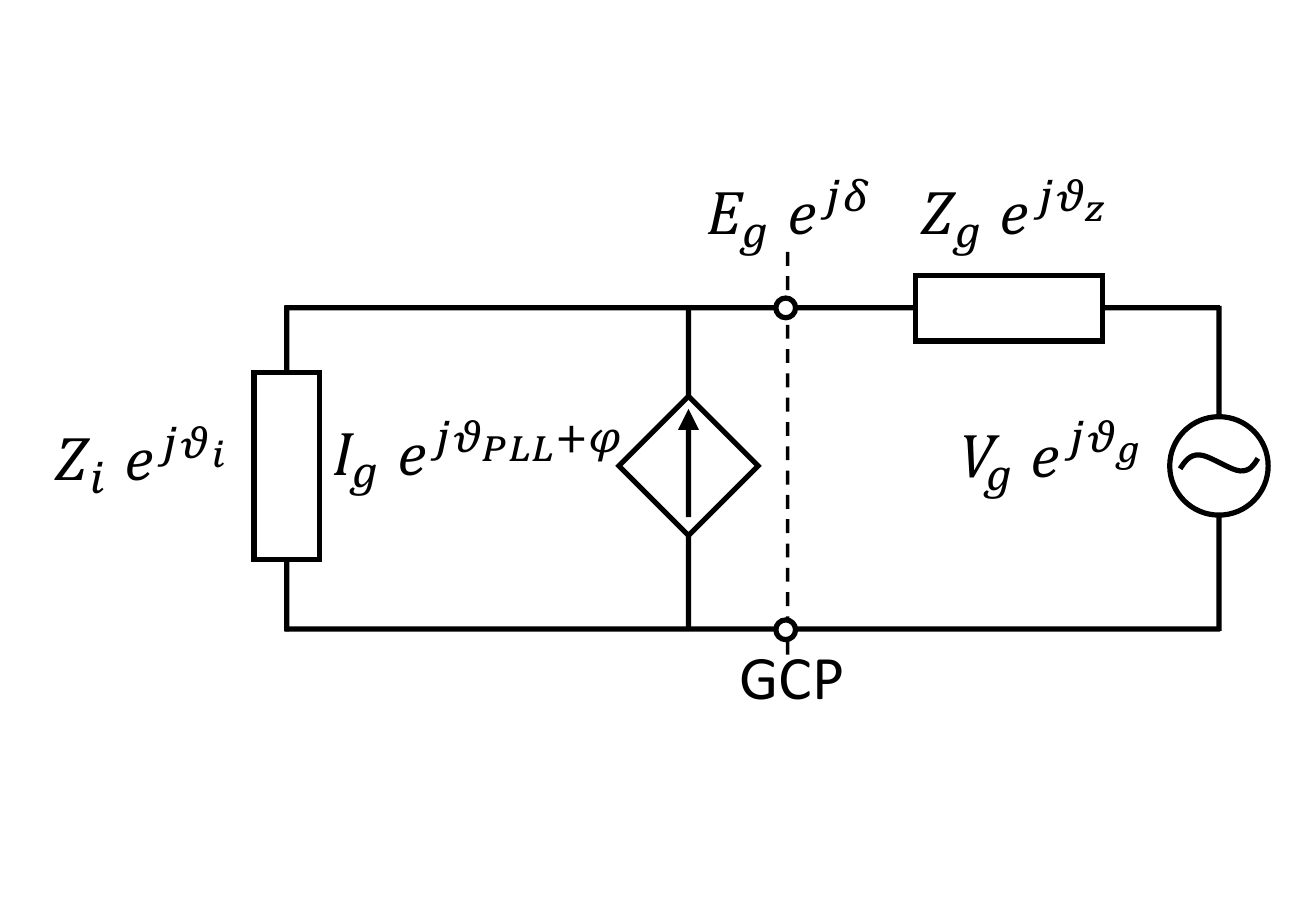}}
\subfloat[ ]{\includegraphics[width=0.5\columnwidth]{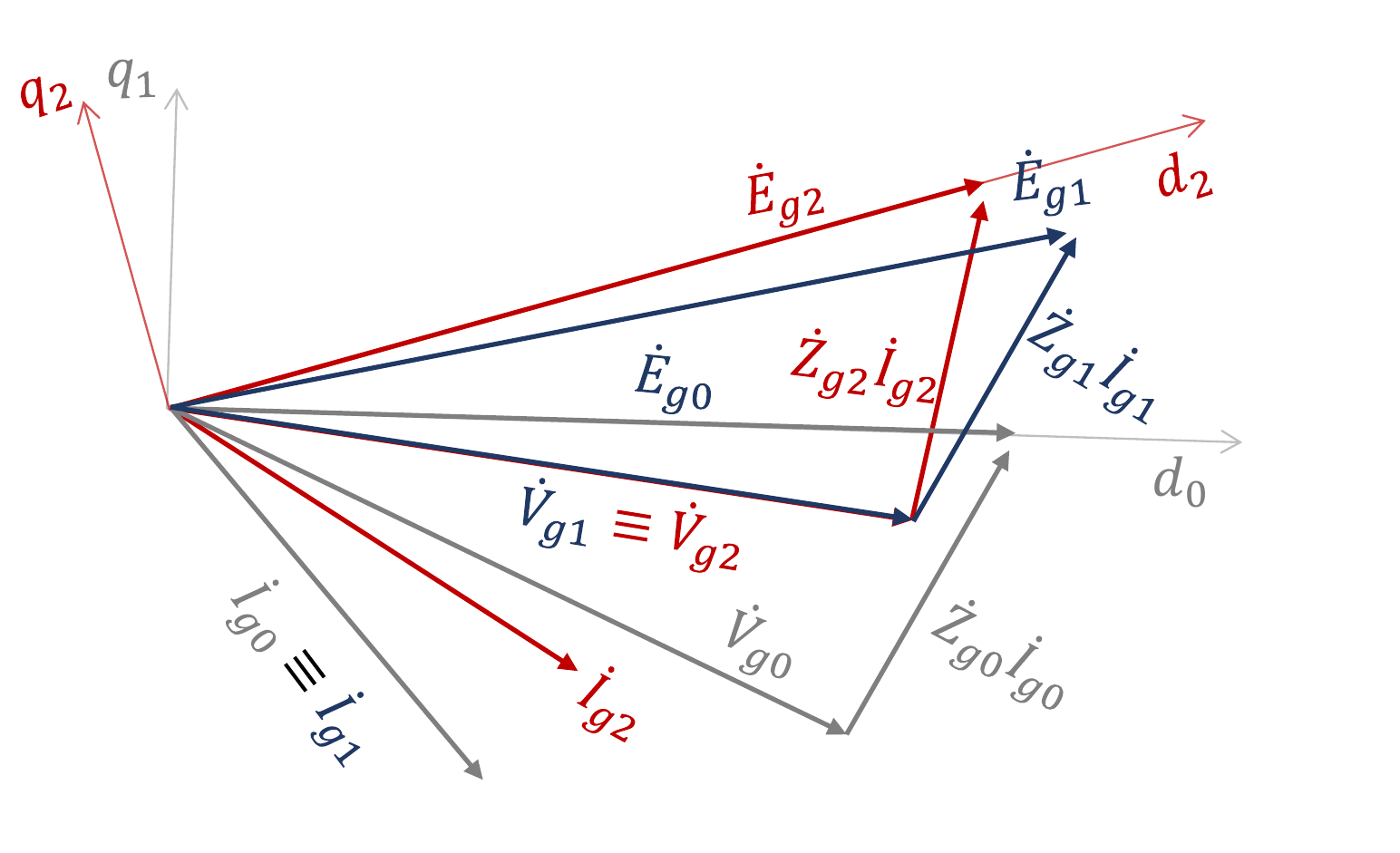}}
\caption{Phasor representation (a) and diagram (b) of the interface between the converter and the grid. Example of vector movements during a grid transient.}
\label{fig:diagramma_e_vettori}
\end{figure}

Let us consider the phasor representation in Fig.~\ref{fig:diagramma_e_vettori}-b. The initial condition (grey vectors) is described by the following equation:
\begin{align}      
    E_{g0} e^{j\delta_0} = I_{g0} Z_{g0} e^{j (\vartheta_{pll0} + \varphi + \vartheta_z)} + V_{g0} e^{j \vartheta_{g0}}. \label{eq:phasor0}
\end{align}
When the grid goes through a transient, the voltage phasor $\dot{V}_{g}$ is shifted by an angle $\Delta \vartheta_{g}$. The converter current phasor remains constant, since it is synchronised with the \ac{pll} reference frame (blue vectors):
\begin{align}      
    E_{g1} e^{j(\delta_0+\Delta\delta)} = I_{g0} Z_{g0} e^{j( \vartheta_{pll0} + \varphi + \vartheta_z)} + V_g e^{j (\vartheta_{g0} +\Delta\vartheta_{g})}.
\end{align}
After a while, the \ac{pll} performs the new synchronisation (red vectors),
\begin{align}      
    E_{g2} e^{j(\delta_0+\Delta\delta+\Delta\vartheta_{pll})} = & I_{g0} Z_{g0} e^{j( \vartheta_{pll0}+\Delta\vartheta_{pll} + \varphi + \vartheta_z)}+ \nonumber \\& + V_{g0} e^{j( \vartheta_{g0} +\Delta\vartheta_{g})},
\end{align}
After the transient, when the frequency is restored, the controller of the converter will act in order to realise its power set-points, as it was before the transient. Therefore, the red phasor diagram should be equal to the grey one, but rotated by $\Delta \vartheta_g$. It can be demonstrated that this is true only if  $E_{g0}=E_{g2}$  and $\Delta\delta = \Delta\vartheta_{g}-\Delta\vartheta_{pll}$.
Please note that this is true with the hypothesis of $V_g$ constant, the same that allows the decoupling between \emph{P--f} and \emph{v--Q} regulation.

 \section{Parameters of the real--time model} \label{appendix:B}
The converter parameters used for both the real--time model and the linear ones are reported in table~\ref{tab:ParamConv1} and \ref{tab:ParamConv2}.
\begin{table}[t]
		\centering
		\caption{Power converter ratings.}
		\label{tab:ParamConv1}
		 \renewcommand{\arraystretch}{1.1}
		 \begin{tabular}{p{1.5cm} p{4.6cm} p{2cm}}
		 \hline \hline
		      & Description             & Value \\\hline 
		 \multicolumn{3}{c}{ \textit{Converter ratings}}\\\hline \hline 
		  $P_n$        & Nominal Power           & \SI{1000}{MW} \\
		  $\cos{\phi_n}$& Nominal power factor    & 0.95\\
		  $V_n$        & Nominal Voltage AC      & \SI{22}{\kilo\volt}\\
		  $V_{n,dc}$   & Nominal Voltage DC      &\SI{44}{\kilo\volt}\\
		  $C_{dc}$     & DC capacitance         & \SI{453.17}{\micro\farad}\\
		  $f_{n}$      & Nominal frequency      & \SI{50}{\hertz}\\		
 		  $f_{s}$ & Switching frequency      & \SI{1.35}{\kilo\hertz}\\
 		  \hline
		  $R_f$        & LC filter resistance   & \SI{0.005}{\pu}\\
		  $L_f$        & LC filter inductance   & \SI{0.200}{\pu}\\
		  $C_f$        & LC filter capacitance   & \SI{0.066}{\pu}\\ 
		  		 \hline \hline
        \end{tabular}
\end{table}

\begin{table}[t]
		\centering
		\caption{Power converter control parameters.}
		\label{tab:ParamConv2}
		 \renewcommand{\arraystretch}{1.02}
		 \begin{tabular}{p{1.5cm} p{6cm} p{2.2cm}}
		 \hline \hline
		      & Description             & Value \\\hline 
		  \multicolumn{3}{c}{ \textit{Device level control}}\\\hline \hline 
		  $k_{P,c}$    & Proportional gain current PI  &  0.73 \\
		  $k_{I,c}$    & Integral gain current PI  &   1.19 \\
		  $k_{P,v}$    & Proportional gain voltage PI  &   0.52 \\
		  $k_{I,v}$    & Integral gain voltage PI  &   1.61 \\ \hline 
		  \multicolumn{3}{c}{\textit{System level control}}\\\hline \hline 
		  \multicolumn{3}{c}{Grid--following converter}\\\hline 
		  $K_p$        & Frequency droop coefficient & \SI{0.02}{\pu} \\
		  $K_q$        & Voltage droop coefficient & \SI{0.02}{\pu} \\	   \hline
		  \multicolumn{3}{c}{Grid--forming converter}\\ \hline 
		  $m_p$        & Active power reverse droop coefficient& \SI{0.02}{\pu} \\ 
		  $\omega_{c}$ & Low pass filter cut-off frequency      & \SI{31.4}{\radian\per\second}\\	
		  $m_q$        & Reactive power reverse droop coefficient & \SI{0.02}{\pu} \\ 			 
		  $k_{RV}$     & Transient Virtual Resistor (TVR) gain  &  0.09 \\ 	
  		  $\omega_{RV}$ & TVR high-pass filter cut-off frequency   & \SI{16.6}{\radian\per\second}\\	
  		  $\sigma_{X/R}$  & Virtual impedance ratio $=X_{vi}/R_{vi}$ & 5 \\ 
  		  $R_{vi}$ & Virtual resistance   & \SI{0.6716}{\pu} \\ 
  		  $T_1$  &  lead-lag filter time constant & \SI{0.033}{\second} \\ 
  		  $T_2$ & lead-lag filter time constant   & \SI{0.011}{\second} \\   		  
  		  \hline
  		  \multicolumn{3}{c}{Phase Locked Loop PLL}\\\hline 
  		  $k_{I,pll}$ & Integral gain \ac{pll} PI   &  180 \\	
  		  $k_{P,pll}$    & Proportional gain \ac{pll} PI  &  3800 \\
  		  $\omega_{f,pll}$ & Second order filter cut--off frequency   & \SI{4}{\hertz}\\		
  		  $\zeta_{f,pll}$        & Second order filter damping ratio  & {0.707}\\	 
		 \hline \hline
        \end{tabular}
\end{table}

\bibliographystyle{elsarticle-num} 

\begin{thebibliography}{10}
\expandafter\ifx\csname url\endcsname\relax
  \def\url#1{\texttt{#1}}\fi
\expandafter\ifx\csname urlprefix\endcsname\relax\def\urlprefix{URL }\fi
\expandafter\ifx\csname href\endcsname\relax
  \def\href#1#2{#2} \def\path#1{#1}\fi

\bibitem{markovic2019understanding}
U.~Markovic, O.~Stanojev, E.~Vrettos, P.~Aristidou, G.~Hug, Understanding
  stability of low-inertia systems, engrXiv (2019).

\bibitem{MilanoPSCC2018}
F.~{Milano}, F.~{Dörfler}, G.~{Hug}, D.~J. {Hill}, G.~{Verbič}, Foundations
  and challenges of low-inertia systems (invited paper), in: 2018 Power Systems
  Computation Conference (PSCC), 2018, pp. 1--25.
\newblock \href {https://doi.org/10.23919/PSCC.2018.8450880}
  {\path{doi:10.23919/PSCC.2018.8450880}}.

\bibitem{PSCC2018}
G.-P. Schiapparelli, E.~Namor, F.~Sossan, R.~Cherkaoui, S.~Massucco,
  M.~Paolone, Quantification of primary frequency control provision from
  battery energy storage systems connected to active distribution networks, in:
  2018 Power Systems Computation Conference (PSCC), 2018, pp. 1--7.
\newblock \href {https://doi.org/10.23919/PSCC.2018.8442554}
  {\path{doi:10.23919/PSCC.2018.8442554}}.

\bibitem{TSE2019}
F.~{Conte}, S.~{Massucco}, G.-P. {Schiapparelli}, F.~{Silvestro}, Day-ahead and
  intra-day planning of integrated {BESS-P}v systems providing frequency
  regulation, IEEE Transactions on Sustainable Energy 11~(3) (2020) 1797--1806.

\bibitem{kundur2004definition}
P.~Kundur, J.~Paserba, V.~Ajjarapu, G.~Andersson, A.~Bose, C.~Canizares,
  N.~Hatziargyriou, D.~Hill, A.~Stankovic, C.~Taylor, T.~Van~Cutsem, V.~Vittal,
  Definition and classification of power system stability ieee/cigre joint task
  force on stability terms and definitions, IEEE Transactions on Power Systems
  19~(3) (2004) 1387--1401.
\newblock \href {https://doi.org/10.1109/TPWRS.2004.825981}
  {\path{doi:10.1109/TPWRS.2004.825981}}.

\bibitem{yuan2017}
H.~{Yuan}, X.~{Yuan}, J.~{Hu}, Modeling of grid-connected vscs for power system
  small-signal stability analysis in dc-link voltage control timescale, IEEE
  Transactions on Power Systems 32~(5) (2017) 3981--3991.

\bibitem{rosso2019}
R.~{Rosso}, J.~{Cassoli}, G.~{Buticchi}, S.~{Engelken}, M.~{Liserre}, Robust
  stability analysis of lcl filter based synchronverter under different grid
  conditions, IEEE Transactions on Power Electronics 34~(6) (2019) 5842--5853.

\bibitem{rosso2019a}
R.~{Rosso}, S.~{Engelken}, M.~{Liserre}, Robust stability analysis of
  synchronverters operating in parallel, IEEE Transactions on Power Electronics
  34~(11) (2019) 11309--11319.

\bibitem{goksu2014instability}
{\"O}.~G{\"o}ksu, R.~Teodorescu, C.~L. Bak, F.~Iov, P.~C. Kj{\ae}r, Instability
  of wind turbine converters during current injection to low voltage grid
  faults and pll frequency based stability solution, IEEE Transactions on Power
  Systems 29~(4) (2014) 1683--1691.
\newblock \href {https://doi.org/10.1109/TPWRS.2013.2295261}
  {\path{doi:10.1109/TPWRS.2013.2295261}}.

\bibitem{rosso2019b}
R.~{Rosso}, M.~{Andresen}, S.~{Engelken}, M.~{Liserre}, Analysis of the
  interaction among power converters through their synchronization mechanism,
  IEEE Transactions on Power Electronics 34~(12) (2019) 12321--12332.

\bibitem{rosso2020}
R.~{Rosso}, S.~{Engelken}, M.~{Liserre}, Robust stability investigation of the
  interactions among grid-forming and grid-following converters, IEEE Journal
  of Emerging and Selected Topics in Power Electronics 8~(2) (2020) 991--1003.

\bibitem{rokrok2019}
E.~{Rokrok}, T.~{Qoria}, A.~{Bruyere}, B.~{Francois}, X.~{Guillaud}, Effect of
  using pll-based grid-forming control on active power dynamics under various
  scr, in: IECON 2019 - 45th Annual Conference of the IEEE Industrial
  Electronics Society, Vol.~1, 2019, pp. 4799--4804.

\bibitem{Wu2018pll_stability}
H.~{Wu}, X.~{Wang}, Transient stability impact of the phase-locked loop on
  grid-connected voltage source converters, in: 2018 International Power
  Electronics Conference (IPEC-Niigata 2018 -ECCE Asia), 2018, pp. 2673--2680.
\newblock \href {https://doi.org/10.23919/IPEC.2018.8507447}
  {\path{doi:10.23919/IPEC.2018.8507447}}.

\bibitem{farrokhabadi2020}
M.~{Farrokhabadi}, C.~A. {Cañizares}, J.~W. {Simpson-Porco}, E.~{Nasr},
  L.~{Fan}, P.~A. {Mendoza-Araya}, R.~{Tonkoski}, U.~{Tamrakar},
  N.~{Hatziargyriou}, D.~{Lagos}, R.~W. {Wies}, M.~{Paolone}, M.~{Liserre},
  L.~{Meegahapola}, M.~{Kabalan}, A.~H. {Hajimiragha}, D.~{Peralta}, M.~A.
  {Elizondo}, K.~P. {Schneider}, F.~K. {Tuffner}, J.~{Reilly}, Microgrid
  stability definitions, analysis, and examples, IEEE Transactions on Power
  Systems 35~(1) (2020) 13--29.

\bibitem{arani2016assessment}
M.~F.~M. Arani, Y.~A.-R.~I. Mohamed, Assessment and enhancement of a full-scale
  pmsg-based wind power generator performance under faults, IEEE Transactions
  on Energy Conversion 31~(2) (2016) 728--739.
\newblock \href {https://doi.org/10.1109/TEC.2016.2526618}
  {\path{doi:10.1109/TEC.2016.2526618}}.

\bibitem{mortazavian2017dynamic}
S.~Mortazavian, Y.~A.-R.~I. Mohamed, Dynamic analysis and improved lvrt
  performance of multiple dg units equipped with grid-support functions under
  unbalanced faults and weak grid conditions, IEEE Transactions on Power
  Electronics 33~(10) (2017) 9017--9032.
\newblock \href {https://doi.org/10.1109/TPEL.2017.2784435}
  {\path{doi:10.1109/TPEL.2017.2784435}}.

\bibitem{shabestary2016analytical}
M.~M. Shabestary, Y.~A.-R.~I. Mohamed, An analytical method to obtain maximum
  allowable grid support by using grid-connected converters, IEEE Transactions
  on Sustainable Energy 7~(4) (2016) 1558--1571.
\newblock \href {https://doi.org/10.1109/TSTE.2016.2569022}
  {\path{doi:10.1109/TSTE.2016.2569022}}.

\bibitem{zhang2016frequency}
W.~Zhang, D.~Remon, P.~Rodriguez, Frequency support characteristics of
  grid-interactive power converters based on the synchronous power controller,
  IET Renewable Power Generation 11~(4) (2016) 470--479.

\bibitem{hammad2017effective}
E.~Hammad, A.~Farraj, D.~Kundur, On effective virtual inertia of storage-based
  distributed control for transient stability, IEEE Transactions on Smart Grid
  10~(1) (2017) 327--336.
\newblock \href {https://doi.org/10.1109/TSG.2017.2738633}
  {\path{doi:10.1109/TSG.2017.2738633}}.

\bibitem{darco2015}
S.~{D’Arco}, J.~A. {Suul}, O.~B. {Fosso}, A virtual synchronous machine
  implementation for distributed control of power converters in smartgrids,
  Electric Power Systems Research 122 (2015) 180--197.

\bibitem{rocabert2012control}
J.~Rocabert, A.~Luna, F.~Blaabjerg, P.~Rodriguez, Control of power converters
  in {AC} microgrids, IEEE transactions on power electronics 27~(11) (2012)
  4734--4749.

\bibitem{green2007control}
T.~C. Green, M.~Prodanovi{\'c}, Control of inverter-based micro-grids, Electric
  power systems research 77~(9) (2007) 1204--1213.

\bibitem{paquette2014}
A.~D. {Paquette}, M.~J. {Reno}, R.~G. {Harley}, D.~M. {Divan}, Sharing
  transient loads : Causes of unequal transient load sharing in islanded
  microgrid operation, IEEE Industry Applications Magazine 20~(2) (2014)
  23--34.

\bibitem{matevosyan2019}
J.~{Matevosyan}, B.~{Badrzadeh}, T.~{Prevost}, E.~{Quitmann},
  D.~{Ramasubramanian}, H.~{Urdal}, S.~{Achilles}, J.~{MacDowell}, S.~H.
  {Huang}, V.~{Vital}, J.~{O’Sullivan}, R.~{Quint}, Grid-forming inverters:
  Are they the key for high renewable penetration?, IEEE Power and Energy
  Magazine 17~(6) (2019) 89--98.

\bibitem{Wang2018}
Y.~{Wang}, X.~{Wang}, Z.~{Chen}, F.~{Blaabjerg}, Small-signal stability
  analysis of inverter-fed power systems using component connection method,
  IEEE Transactions on Smart Grid 9~(5) (2018) 5301--5310.

\bibitem{migrate3_2}
T.~Qoria, Q.~Cossart, C.~Li, X.~Guillaud, F.~Colas, F.~Gruson, X.~Kestelyn,
  Deliverable 3.2: Local control and simulation tools for large transmission
  systems, in: MIGRATE project, https://www.h2020-migrate.eu/downloads.html,
  2018.

\bibitem{wen2013influence}
B.~Wen, D.~Boroyevich, P.~Mattavelli, Z.~Shen, R.~Burgos, Influence of
  phase-locked loop on input admittance of three-phase voltage-source
  converters, in: 2013 Twenty-Eighth Annual IEEE Applied Power Electronics
  Conference and Exposition (APEC), 2013, pp. 897--904.

\bibitem{qoria2019power}
T.~Qoria, T.~Prevost, G.~Denis, F.~Gruson, F.~Colas, X.~Guillaud, Power
  converters classification and characterization in power transmission systems,
  in: 21st European Conference on Power Electronics and Applications (EPE '19
  ECCE Europe), Genova, Italy, 2019.

\bibitem{sun2011impedance}
J.~Sun, Impedance-based stability criterion for grid-connected inverters, IEEE
  transactions on power electronics 26~(11) (2011) 3075--3078.

\bibitem{wang2014modeling}
X.~Wang, F.~Blaabjerg, W.~Wu, Modeling and analysis of harmonic stability in an
  {AC} power-electronics-based power system, {IEEE Transactions on Power
  Electronics} 29~(12) (2014) 6421--6432.

\bibitem{bottrell2013}
N.~{Bottrell}, M.~{Prodanovic}, T.~C. {Green}, Dynamic stability of a microgrid
  with an active load, IEEE Transactions on Power Electronics 28~(11) (2013)
  5107--5119.

\bibitem{guo2014distributed}
F.~Guo, C.~Wen, J.~Mao, Y.-D. Song, Distributed secondary voltage and frequency
  restoration control of droop-controlled inverter-based microgrids, {IEEE
  Transactions on Industrial Electronics} 62~(7) (2014) 4355--4364.

\bibitem{kerdphol2019self}
T.~Kerdphol, M.~Watanabe, K.~Hongesombut, Y.~Mitani, Self-adaptive virtual
  inertia control-based fuzzy logic to improve frequency stability of microgrid
  with high renewable penetration, IEEE Access 7 (2019) 76071--76083.

\bibitem{bevrani2014robust}
H.~Bevrani, Robust power system frequency control, Springer, 2014.

\bibitem{Gu2018}
Y.~{Gu}, N.~{Bottrell}, T.~C. {Green}, Reduced-order models for representing
  converters in power system studies, IEEE Transactions on Power Electronics
  33~(4) (2018) 3644--3654.

\bibitem{timbus2009evaluation}
A.~Timbus, M.~Liserre, R.~Teodorescu, P.~Rodriguez, F.~Blaabjerg, Evaluation of
  current controllers for distributed power generation systems, IEEE
  Transactions on Power Electronics 24~(3) (2009) 654--664.
\newblock \href {https://doi.org/10.1109/TPEL.2009.2012527}
  {\path{doi:10.1109/TPEL.2009.2012527}}.

\bibitem{kundur1994power}
P.~Kundur, N.~J. Balu, M.~G. Lauby, Power system stability and control, Vol.~7,
  McGraw-hill New York, 1994.

\bibitem{ieee1992hydraulic}
{IEEE Working Group and others}, Hydraulic turbine and turbine control models
  for system dynamic studies, IEEE Transactions on Power Systems 7~(1) (1992)
  167--179.
\newblock \href {https://doi.org/10.1109/59.141700}
  {\path{doi:10.1109/59.141700}}.

\bibitem{saccomanno2003electric}
F.~Saccomanno, Electric power systems: analysis and control,
  Wiley-Interscience, 2003.

\bibitem{ramos2015benchmark}
R.~Ramos, I.~Hiskens, et~al., Benchmark systems for small-signal stability
  analysis and control, Tech. Rep. PES-TR18, IEEE Power and Energy Society
  (August 2015).

\bibitem{dervivskadic2018under}
A.~Dervi{\v{s}}kadi{\'c}, Y.~Zuo, G.~Frigo, M.~Paolone, Under frequency load
  shedding based on pmu estimates of frequency and rocof, in: 2018 IEEE PES
  Innovative Smart Grid Technologies Conference Europe (ISGT-Europe), IEEE,
  2018, pp. 1--6.

\bibitem{zuo2018dispatch}
Y.~Zuo, F.~Sossan, M.~Bozorg, M.~Paolone, Dispatch and primary frequency
  control with electrochemical storage: a system-wise verification, in: 2018
  IEEE PES Innovative Smart Grid Technologies Conference Europe (ISGT-Europe),
  IEEE, 2018, pp. 1--6.

\bibitem{githubDESLEPFL}
DESL-EPFL.
\newblock \href{https://github.com/DESL-EPFL}{{Distributed Electrical Systems
  Laboratory (DESL), Ecole Polytechnique F{\'e}d{\'e}rale De Lausanne (EPFL)}}
  [online] (2019) [cited 2019-08-07].

\end{thebibliography}





\end{document}